\documentclass{aa}  

\usepackage{footnote}
\usepackage{graphicx}
\usepackage{hyperref}
\usepackage{lineno}
\usepackage{appendix}
\usepackage{color}
\usepackage{soul}
\usepackage{natbib}
\usepackage{amsmath}
\usepackage{xspace}
\usepackage{enumitem}
\usepackage{amssymb}
\usepackage{xifthen}
\usepackage[normalem]{ulem}
\usepackage{amsfonts}
\usepackage{multirow}
\usepackage{dsfont}
\usepackage{txfonts}
\newcommand{\code}[1]{\texttt{#1}\xspace}

\begin{document}

   \title{Evidence for multiple nucleosynthetic processes from carbon enhanced metal-poor stars in the Carina dwarf spheroidal galaxy}
\titlerunning{Carbon rich stars in Carina}
\authorrunning{Hansen et al.}


   \author{T. T. Hansen
          \inst{1}
          \and
          J. D. Simon\inst{2}
          \and
          T. S. Li\inst{3}
          \and
          A. Frebel\inst{4}
          \and
          I. Thompson\inst{2}
          \and
          S. Shectman\inst{2}
          }

   \institute{Department of Astronomy, Stockholm University, AlbaNova
University Center, SE-106 91 Stockholm, Sweden \email{terese.hansen@astro.su.se}
         \and
             Observatories of the Carnegie Institution for Science, 813 Santa Barbara St., Pasadena, CA 91101, USA
            \and
            Department of Astronomy and Astrophysics, University of Toronto, 50 St. George Street, Toronto ON, M5S 3H4, Canada
            \and
            Department of Physics \& Kavli Institute for Astrophysics and Space Research, Massachusetts Institute of Technology, Cambridge, MA 02139, USA
             }


 
  \abstract
   {Carbon Enhanced Metal-Poor (CEMP) stars ($\mathrm{[C/Fe]} > 0.7$) are known to exist in large numbers at low metallicity in the Milky Way halo and are important tracers of early Galactic chemical evolution. However, very few such stars have been identified in the classical dwarf spheroidal (dSph) galaxies, and detailed abundances, including neutron-capture element abundances, have only been reported for 13 stars.}
   {We aim to derive detailed abundances of six CEMP stars identified in the Carina dSph and compare the abundances to CEMP stars in other dSph galaxies and the Milky Way halo. This is the largest sample of CEMP stars in a dSph galaxy analysed to date. }
   {1D LTE elemental abundances are derived via equivalent width and spectral synthesis using high-resolution spectra of the six stars, obtained with the MIKE spectrograph at Las Campanas Observatory.}
   {Abundances or upper limits are derived for up to 27 elements from C to Os in the six stars. The analysis reveals one of the stars to be a CEMP-no star with very low neutron-capture element abundances. In contrast, the other five stars all show enhancements in neutron-capture elements in addition to their carbon enhancement, classifying them as CEMP-$s$ and -$r/s$ stars. The six stars have similar $\alpha$ and iron-peak element abundances as other stars in Carina, except for the CEMP-no star, which shows enhancement in Na, Mg, and Si. We explore the absolute carbon abundances ($A(\rm C)$) of CEMP stars in dSph galaxies and find similar behaviour as is seen for Milky Way halo CEMP stars, but highlight that CEMP-$r/s$ stars primarily have very high $A(\rm C)$ values. We also compare the neutron-capture element abundances of the CEMP-$r/s$ stars in our sample to recent $i$-process yields, which provide a good match to the derived abundances.}
   {}


   \maketitle

%

\section{Introduction \label{sec:intro}}
It has long been recognized that a substantial fraction of the metal-poor stars in the Milky Way (MW) halo show moderate to large enhancements in carbon. These stars, which have $\mathrm{[C/Fe]} > 0.7$, are labelled CEMP (Carbon Enhanced Metal-Poor) stars \citep{aoki2007,beers2005}. \cite{placco2014} found that 20\% of halo stars with $\mathrm{[Fe/H]} < -2$ are CEMP stars, with the fraction increasing with decreasing metallicity. Hence this abundance signature is pervasive in the early universe, and these stars are important tools for studying early Galactic chemical evolution. The CEMP stars are divided into sub-categories depending on their neutron-capture element abundances: CEMP-no stars, which show no enhancement in neutron-capture elements ($\mathrm{[Ba/Fe]} < 0.0$), and the CEMP-$r$,-$s$, and -$r/s$ stars, which apart from the carbon also show enhancements in rapid neutron capture $(r-)$process (CEMP-$r$: $\mathrm{[Eu/Fe]} > 1.0$), slow neutron-capture $(s-)$process elements (CEMP-$s$: $\mathrm{[Ba/Fe]} > 1.0$, $\mathrm{[Ba/Eu]} > 0.5$), 
or a combination of both (CEMP-$r/s$: $0.0 < \mathrm{[Ba/Eu]} < 0.5$) \citep{beers2005}. 

Radial-velocity monitoring of CEMP-no and CEMP-$s$ stars has revealed that the majority of CEMP-$s$ stars are part of binary systems while the CEMP-no stars are mostly found to be single stars \citep{lucatello2005,starkenburg2014,hansen2016a,hansen2016b}. These results support the proposed formation scenario for the CEMP-$s$ stars, namely that the peculiar abundance pattern of these stars is the result of mass transfer from a companion which has passed through the asymptotic giant branch (AGB) phase, thus producing C and $s$-process elements. The non-binary result for the CEMP-no stars, however, points to these stars likely being second-generation stars with abundance patterns that are fingerprints of nucleosynthesis processes having taken place in the first generations of stars. Limited radial-velocity monitoring of CEMP-$r/s$ stars suggests that these, like the CEMP-$s$ stars, are also predominantly found in binary systems \citep{hansen2016b}. The CEMP-no stars thus provide essential information on the nucleosynthesis of the first stars to form in the Universe, while the CEMP-$s$ and likely $r/s$ stars can be used to gain information about neutron-capture processes in the early Universe as well as mass transfer across binary systems.

Although the abundances of CEMP stars in the MW halo have been the focus of multiple studies \citep[e.g.,][]{aoki2007,norris2013,hansen2019}, initial spectroscopic exploration of the dSph galaxies has not revealed CEMP stars at similar rates as has been found in the halo \citep{simon15,salvadori2015}. This result is puzzling, as the MW halo is believed to be built via the accretion of surrounding smaller satellite systems \citep{searle1978}, a conclusion that is strongly supported by the numerous stellar streams discovered in the MW halo in recent large photometric surveys and, in particular, in the data from the Gaia satellite \citep[e.g.,][]{malhan2018,li2019}. One possible explanation for the apparent rarity of CEMP stars in dwarf galaxies could be the difference in metallicity selection. Many of the halo studies have targeted stars with $\mathrm{[Fe/H]} < -2.5$, whereas relatively few stars in dSphs with $\mathrm{[Fe/H]} < -2.5$ have been studied. Indeed a dedicated search at low metallicity has resulted in 22 CEMP stars being detected in Sculptor, uncovering a CEMP fraction at low metallicity similar to that of the MW halo \citep{chiti2018}. However, similar dedicated searches for CEMP stars in other dSph galaxies are still lacking.

In addition, to the low number of CEMP stars currently discovered in dSph galaxies, even fewer have been subject to high-resolution abundance studies. Thus, very little is known about the CEMP population in our dwarf galaxy neighbours and whether it is similar to what we see in the MW halo. This paper presents the detailed abundance analysis of six CEMP stars in the Carina dSph galaxy. This is the largest sample of CEMP stars analysed in a dSph galaxy to date. In Section \ref{sec:obs}, we present the observational data for the six stars. The abundance analysis is outlined in Section \ref{sec:analysis}, and the results presented in Section \ref{sec:results}. Section \ref{sec:discuss} provides a discussion of our results, and a summary is given in Section \ref{sec:summary}

\section{Observations \label{sec:obs}}
The sample stars were first observed as part of a larger search for metal-poor stars in Carina and other dSphs carried out with the IMACS spectrograph \citep{dressler2011} on the Magellan/Baade telescope at Las Campanas Observatory. In the first stage, low-resolution spectra of a total of $\sim1300$ candidate Carina stars were obtained with four overlapping slit masks, similar to the process described by \citet{chiti2018} for Sculptor. Spectroscopic targets were selected from the Carina photometric catalogues available at the Canadian Astronomy Data Centre\footnote{https://www.canfar.net/storage/list/STETSON/homogeneous/Carina/} \citep{stetson2000,stetson2005}, which are updated versions of the data originally published by \citet{smeckerhane1994}.  The IMACS spectra were taken through a narrow-band filter targeting the Ca~K line to maximize the multiplexing advantage offered by the IMACS field of view.  Stars with weak Ca~K lines were selected for follow-up observations at medium spectral resolution with the MagE spectrograph  \citep{marshall2008} on the Magellan/Clay telescope. The MagE spectra cover a wavelength range from 3100\AA\ to 1 micron and were obtained during four observing runs in 2012 February, 2012 December, 2013 February, and 2013 September using the 0\farcs7 or 1\farcs0 slits (depending on the seeing) and $1\times1$ binning yielding a resolving power of $R\sim4100$.

\begin{figure}[hbt!]
\centering
\includegraphics[width=\linewidth]{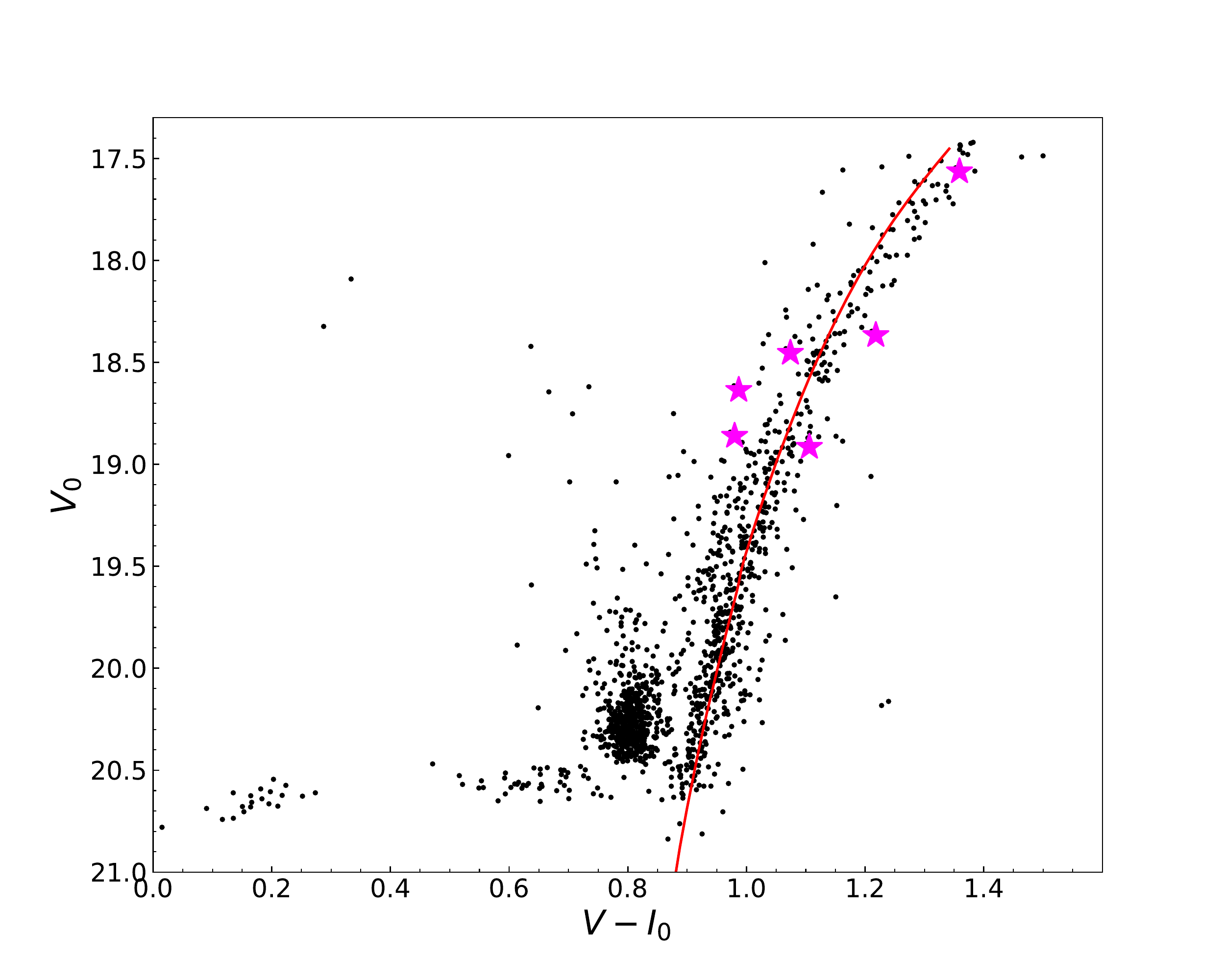}
\caption{Color-magnitude diagram for Carina. Black dots are stars identified in \citet{fabrizio2016}, and pink stars are the six stars analysed in this paper. Red curve show a Dartmouth isochrone \citep{dotter2008} with $\mathrm{[Fe/H]} = -1.75$, age = 12 Gyr, and $\mathrm{[\alpha/Fe]=0.4}$ shifted to the distance of Carina using the distance moduli (m-M)$_V$=20.05 mag and (m-M)$_I$=19.98 mag from \citet{mighell1997}. \label{fig:cmd}}
\end{figure}

Out of a total of 25 Carina stars observed with MagE, eight exhibited strong C-H features, and we obtained high-resolution spectroscopy of six of those with the MIKE spectrograph \citep{bernstein2003} at the Magellan/Clay telescope over two runs in 2016 December and 2019 April. The spectra cover a wavelength range of 3350-5000 \AA\ in the blue and 4900-9500 \AA\ in the red and were obtained using a combination of the $1\arcsec\times5\arcsec$ and $0\farcs7\times5\arcsec$ slits and $2\times2$ binning yielding typical resolving powers of 35,000 in the blue and 28,000 in the red for the $0\farcs7\times5\arcsec$ slit and 28,000 in the blue and 22,000 in the red for the $1.0\farcs\times5\farcs$ slit. The data were reduced with the CarPy MIKE pipeline \citep{kelson2003}. Table \ref{tab:obslog} lists our targets (IDs from \citet{fabrizio2016}), total exposure times, and signal-to-noise ratios (S/N) of the reduced and co-added spectra. In addition, Figure \ref{fig:cmd} shows a colour-magnitude diagram of Carina stars using photometry from \citet{fabrizio2016} with pink star symbols marking the six stars analysed in this work. All photometry has been corrected for reddening using the dustmaps from \citet{schlafly2011}. Overplotted in red is a Dartmouth isochrone \citep{dotter2008} with $\mathrm{[Fe/H]} = -1.75$, age = 12 Gyr, and $\mathrm{[\alpha/Fe]=0.4}$, shifted to the distance of Carina using the distance moduli (m-M)$_V$=20.05 mag and (m-M)$_I$=19.98 mag from \citet{mighell1997}, describing the intermediate red giant branch population identified in \citet{fabrizio2016}.

\begin{table*}
\caption{Observation Log for MIKE Data.}
\label{tab:obslog}
    \centering
    \begin{tabular}{lrrcrr}
    \hline\hline
    Stellar ID & RA & Dec & total exp time & SN @ 4500\AA & SN @ 6500\AA \\
    \hline
75044 & 06 41 08.58 & $-$50 47 50.1 & 3h       &20 &50\\ 
86357 & 06 41 23.27 & $-$51 13 58.7 & 4h 20min &10 &25\\
130624 & 06 42 33.03 & $-$50 53 30.7 & 3h 20min &15 &25\\
125696 & 06 42 23.94 & $-$50 54 03.2 & 2h 10min &12 &22\\
40475 & 06 40 10.95 & $-$51 01 32.6 & 5h 15min &15 &30\\
97508 & 06 41 39.95 & $-$50 42 41.3 & 1h       &7  &12\\
\hline
    \end{tabular}
\end{table*}

\subsection{Radial Velocities}
Heliocentric radial velocities of our target stars were determined via cross-correlation of the spectra. The MIKE spectra were cross-correlated against the standard star HD~122563 ($rv_{hel} = -26.17$ \citep{gaiarv}) employing 20 to 50 orders in each spectrum depending on the S/N. 

We also measured velocities from the MagE spectra from cross-correlation with the spectrum of 75044, which has the highest S/N, using 3 to 8 orders. The heliocentric radial velocity for 75044 was determined to be 225.7$\pm$1.9 km\,s$^{-1}$ using the \code{IRAF} task rvidlines identifying lines of Na (5889,5895), Mg (5183), and Ba (4934, 5853, 6141). Our final radial velocities are listed in Table \ref{tab:rv}, along with values from the literature. For the MIKE spectra, we list the standard deviation of radial velocities from the individual orders as the error, while for the MagE spectra, we list the 1.9 km\,s$^{-1}$ velocity error on our template star and the standard deviation of radial velocities from the individual orders added in quadrature. 

\begin{table}
\caption{Heliocentric Radial Velocities for the Sample Stars.}
\label{tab:rv}
    \centering
    \resizebox{\columnwidth}{!}{
    \begin{tabular}{llc|llc}
\hline\hline
HJD & V$_{helio}$(err) & ref & HJD & V$_{helio}$(err) & ref\\
 & [km\,s$^{-1}$] & & & [km\,s$^{-1}$] & \\
\hline
\multicolumn{3}{c}{75044} &\multicolumn{3}{c}{86357}\\
\hline
2454122.587 & 222.2(5.7)   & 1 &  \dots     & 235.6(11.6)& 3 \\ 
2456280.828 & 225.7(1.9)   & 2 &  2456280.814 & 225.1(2.1) & 2 \\
2457744.633 & 225.5(1.1)   & 2 &  2457744.700 & 227.6(1.3) & 2 \\
2458599.518 & 225.5(1.0)   & 2 &  2457751.599 & 229.5(1.2) & 2 \\
\hline
\multicolumn{3}{c}{130624}&\multicolumn{3}{c}{125696}\\
\hline
\dots       & 210.2(1.6)& 3& \dots         & 212.3(8.1)&3\\
2454123.555 & 208.2(6.6)& 1&  2454122.587 & 214.3(8.1)&1\\
2455982.683 & 211.7(3.8)& 2&  2456332.566 & 221.8(2.8)&2\\
2457751.698 & 218.5(0.7)& 2&  2457751.806 & 228.0(0.8)&2\\
2457752.842 & 219.8(1.0)& 2&              &           & \\
\hline
\multicolumn{3}{c}{40475}& \multicolumn{3}{c}{97508}\\
\hline
\dots       & 238.4(10.3) &3 &  \dots        & 231.8(14.2)&3 \\
\dots       & 231.9(1.6)  &4 &  2456331.834 & 229.5(2.2) &2 \\
2456329.743 & 233.6(2.0)  &2 &  2457752.803 & 232.3(1.0) &2 \\
2457752.656 & 236.2(0.7)  &2 &              &            &  \\
\hline
    \end{tabular}}
    \tablebib{(1) \cite{walker2009}; (2) This work; (3) \cite{fabrizio2016}; (4) \cite{munoz2006}}
    
\end{table}

Inspection of Table \ref{tab:rv} reveals that two of the stars, 130624 and 125696, exhibit clear radial-velocity variations consistent with being part of binary systems. We used a $\chi^2$ test of the probability $p$ that each star's velocity sequence from Table \ref{tab:rv} is consistent with a constant velocity to show that $p \ll 0.05$ for both stars. For 86357 and 40475, the measurements also suggest binary motion, but the uncertainties on some of the measurements are too large to reach firm conclusions. 
The $\chi^2$ probabilities for these stars are $p = 0.27$ and $p = 0.07$, respectively, so more velocity measurements are needed. Finally, our measurements for 97508 ($p = 0.51$) and 75044 ($p = 0.95$) agree with literature values, suggesting no binary motion. However, for 75044 \cite{susmitha2017} measured radial velocities differing by 30~km\,s$^{-1}$ for two epochs for this star\footnote{Exact radial velocities are not listed in \cite{susmitha2017}.}. Thus this star is also likely a binary.

\section{Stellar parameter determination and abundance analysis\label{sec:analysis}} 
Effective temperatures ($T_{\rm eff}$) for the stars have been determined from $V-I$, $V-J$, $V-H$, $V-K$, and $J-K$ colours using the empirical relations from \cite{casagrande2010}. $V$ and $I$ magnitudes are from \citet{fabrizio2016}, and 2MASS $JHK$ magnitudes were taken from \citet{cutri2003}. All magnitudes were corrected for reddening using \cite{schlafly2011}. The available photometry varies for the six stars. For 75044, 86357, and 130624, two or more colours, including the $V-I$ index, were used, and the final temperature was computed as an average of the individual temperatures. On the other hand, for 125696, 40475, and 97508, only the $V-I$ index could be used. A summary of the photometry and temperature determinations for all stars is listed in Table \ref{tab:photo} in the Appendix. $B$ magnitudes are also available for the stars \citep{fabrizio2016}. However, these are known to be affected by the strong CH bands in CEMP stars, so we disregard the $B-V$ colours when determining the temperatures of our stars. Following the determination of $T_{\rm eff}$, surface gravity ($\log g$), metallicity ($\mathrm{[Fe/H]}$), and microturbulence ($\xi$) were determined spectroscopically from equivalent width (EW) measurements of \ion{Fe}{i} and \ion{Fe}{ii} lines. Atomic data and individual abundances for the Fe lines used in the analysis are listed in Table \ref{tab:lines}. The final stellar parameters are presented in Table \ref{tab:param}. For the one star where we have temperature estimates from multiple colours (see Table \ref{tab:photo} in the Appendix), we find a standard deviation of 65~K. However, since, for the majority of stars, fewer colours are used to determine the temperature, we estimate uncertainties on our stellar parameters of 150~K on T$_{eff}$, 0.3~dex on $\log g$ and 0.2~km\,s$^{-1}$ on $\xi$. For the metallicity, the standard deviation of the \ion{Fe}{ii} line abundances is listed in Table \ref{tab:param}. The stars 125696, 40475, and 97508 were also analysed by \citet{koch2006} who finds $\mathrm{[Fe/H]} = -2.33\pm0.43$, $-2.66\pm0.09$, and $-2.22\pm0.10$ respectively, which agrees within uncertainties with the values derived in this work \footnote{Note that no radial velocities are presented in \citet{koch2006}, however, the same data are used to determine the velocities presented in \citet{fabrizio2016}, which are included in Table \ref{tab:rv}}.

\begin{table}[]
\caption{Stellar Parameters of our target stars.}
\label{tab:param}
    \centering
    \resizebox{\columnwidth}{!}{
    \begin{tabular}{lrrrr}
    \hline\hline
Stellar ID & $T_{\rm eff}$ & $\log g$ & $\mathrm{[Fe/H]}$ & $\xi$ \\
  & ($\pm$150~K) & ($\pm$0.3~dex) &   & ($\pm$0.2~km\,s$^{-1}$)\\ 
  \hline
75044 & 4140 & 0.56 & $-$2.63$\pm$0.11 & 2.68 \\  
86357 & 4329 & 0.97 & $-$2.87$\pm$0.28 & 2.76 \\  
130624 & 4622 & 0.03 & $-$2.02$\pm$0.21 & 2.14 \\  
125696 & 4868 & 1.37 & $-$1.74$\pm$0.22 & 2.04 \\  
40475 & 4546 & 1.16 & $-$2.73$\pm$0.16 & 2.38 \\   
97508 & 4866 & 1.76 & $-$2.06$\pm$0.21 & 2.36 \\   
\hline
    \end{tabular}}
\end{table}

Abundances for the six stars in Carina have been derived via EW analysis as well as spectral synthesis using the analysis code 
\code{SMHR}\footnote{\href{https://github.com/andycasey/smhr}{https://github.com/andycasey/smhr}} to run the 2017 version of the 1D radiative transfer code \code{MOOG}\footnote{\href{https://github.com/alexji/moog17scat}{https://github.com/alexji/moog17scat}}\citep{sneden1973,sobeck2011}, assuming local thermodynamical equilibrium. $\alpha$-enhanced ($\mathrm{[\alpha/Fe] = +0.4}$) \code{ATLAS9} model atmospheres \citep{castelli2003} and line lists generated from \code{linemake}\footnote{\href{https://github.com/vmplacco/linemake}{https://github.com/vmplacco/linemake}} \citep{placco2021} updated with C-H lines from \cite{masseron2014}\footnote{\href{https://nextcloud.lupm.in2p3.fr/s/r8pXijD39YLzw5T?path=\%2FCH}{https://nextcloud.lupm.in2p3.fr/s/r8pXijD39YLzw5T?path=\%2FCH}} were used as input, and Solar abundances were taken from \citet{asplund2009}. The synthesis also includes isotopic and hyperfine structure broadening, where applicable, employing the $s$-process isotope ratios from \citet{sneden2008}. Table \ref{tab:lines} in the appendix lists the atomic data for the lines used in the analysis, along with the measured EW and abundance of each line. To determine our final abundances and associated uncertainties, we follow the procedure outlined in \cite{ji2020b}, which computes a mean abundance weighted by the S/N of the individual lines analysed, and the uncertainty calculation includes both systematic and statistical uncertainties as well as co-variance terms between stellar parameters.

\section{Results\label{sec:results}}

\begin{figure*}[hbt!]
\centering
\includegraphics[width=\linewidth]{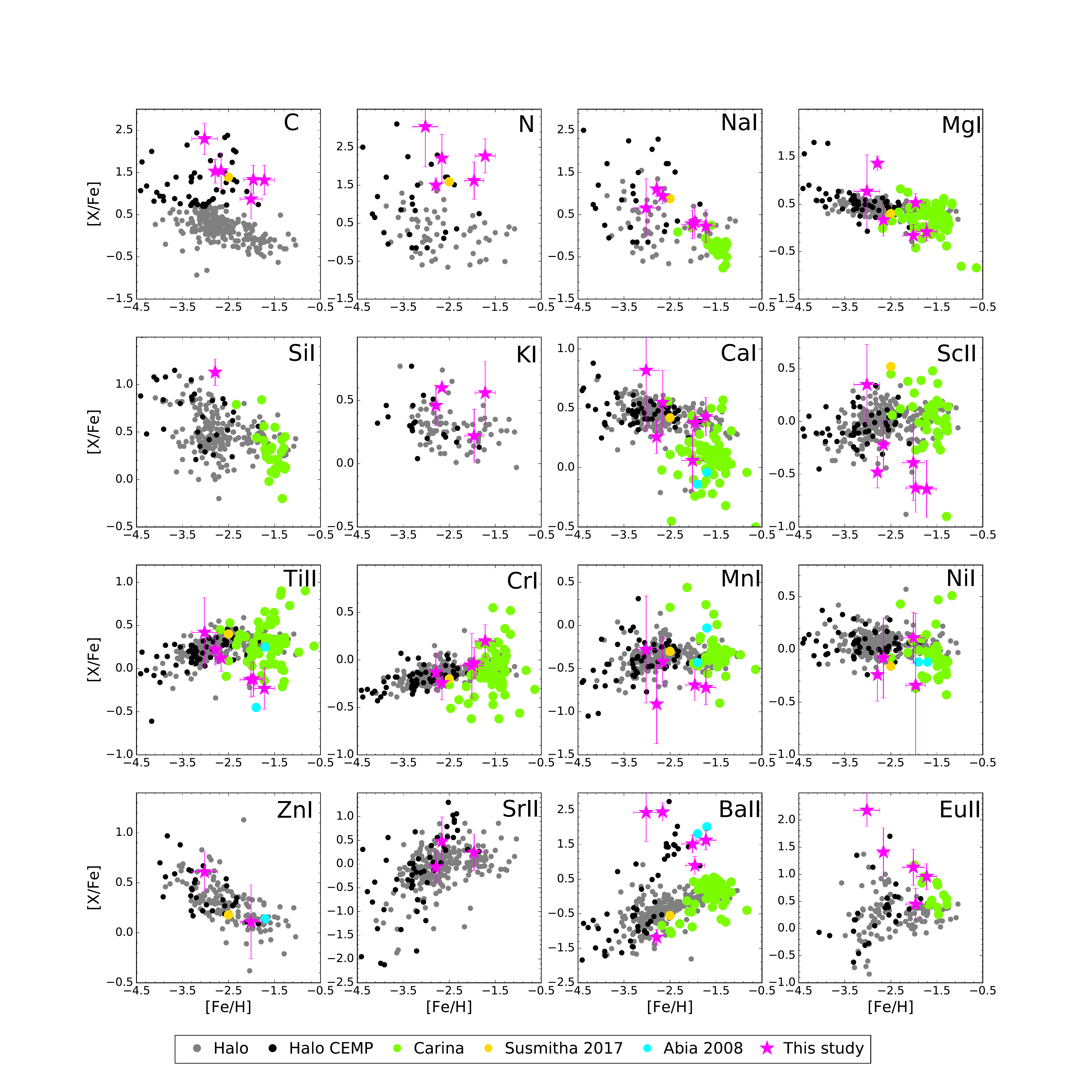}
\caption{Abundances for the six stars analysed here (pink stars) compared to abundances for carbon-normal Carina stars (filled green circles), CEMP stars in Carina (yellow and cyan circles), carbon-normal MW halo stars (small grey circles), and CEMP MW halo stars (small black circles). All literature abundances have been converted to the \cite{asplund2009} Solar abundance scale, and all C abundances corrected for evolutionary effects following \citet{placco2014}.\label{fig:tot}}
\end{figure*}

We derive abundances for up to 14 light elements from C to Zn and 13 neutron-capture elements for our six Carina stars. Final abundances are listed in Tables \ref{tab:abun1} and \ref{tab:abun2}, while Table \ref{tab:err} lists the abundance uncertainties due to stellar parameter uncertainties ($\Delta_X$) and the systematic uncertainties ($s_X$) for each element in each star. $s_X$ is based on the line to line scatter in the abundances and is included to account for systematic uncertainties in, for example, the atomic data or continuum placement. Abundances for selected elements are plotted in Figure \ref{fig:tot} along with abundances reported in the literature for carbon-normal and CEMP stars in Carina and the MW halo. The abundances for Carina stars (green filled circles) are mainly from \citet{norris2017}, which includes a re-analysis of the data from \citet{shetrone2003}, \citet{venn2012}, and \citet{lemasle2012}. We have also included abundances for additional stars analysed by \citet{koch2008} and \citet{fabrizio2015}. For the Carina CEMP stars, the data is taken from \citet{abia2008} (cyan circles) and \citet{susmitha2017} (yellow circles), while the MW halo carbon-normal (small grey circles) and CEMP (small black circles) star abundances are from \citet{roederer2014}. We note that \citet{roederer2014} used spectroscopically derived effective temperatures leading to slightly lower metallicities being derived for the stars compared to photometric temperatures as used in this work. However, \citet{roederer2014} provides homogeneous abundances for a large number of elements, including C, N, K, Zn, and Eu, which are interesting for comparison to the stars analysed here.

\begin{table*}
\caption{Abundance Summary for Stars 75044, 86357, and 130624}
\label{tab:abun1}
    \centering
    \resizebox{\textwidth}{!}{
    \begin{tabular}{lrrrrrrrrrrrrrrrrrr}
    \hline\hline
     & \multicolumn{6}{c}{\textbf{75044}} &\multicolumn{6}{c}{\textbf{86357}} &\multicolumn{6}{c}{\textbf{130624}}\\
 El. & N & $\log\epsilon$ (X) & $\mathrm{[X/H]}$ & $\sigma_{\mathrm{[X/H]}}$ & $\mathrm{[X/Fe]}$ & $\sigma_{\mathrm{[X/Fe]}}$ & N & $\log\epsilon$ (X) & $\mathrm{[X/H]}$ & $\sigma_{\mathrm{[X/H]}}$ & $\mathrm{[X/Fe]}$ & $\sigma_{\mathrm{[X/Fe]}}$ & N & $\log\epsilon$ (X) & $\mathrm{[X/H]}$ & $\sigma_{\mathrm{[X/H]}}$ & $\mathrm{[X/Fe]}$ & $\sigma_{\mathrm{[X/Fe]}}$\\
 &  &  &  & [dex] & & [dex] &  &  &  & [dex] &  & [dex] &  &  &  & [dex] &  & [dex]\\
\hline
%
 C-H        &  1  & $+$6.89 &  $-$1.54 & 0.43 & $+$1.12 & 0.42  &     & \dots &  \dots &\dots&\dots &\dots&  1  & $+$7.59 &  $-$0.84 & 0.35 & $+$1.12 & 0.34  \\  
 C$_2$        &  1  & $+$7.05 &  $-$1.39 & 0.34 & $+$1.27 & 0.34  &  1  & $+$7.46 & $-$0.97 & 0.49 & $+$2.05 & 0.37  &  1  & $+$7.39 &  $-$1.04 & 0.17 & $+$0.92 & 0.16  \\ 
 C$_{corr}$ &     & $+$7.31 &  $-$1.12 & 0.43 & $+$1.54 & 0.42  &     & $+$7.71 & $-$0.72 & 0.49 & $+$2.30 & 0.37  &  1  & $+$7.80 &  $-$0.63 & 0.35 & $+$1.33 & 0.34  \\  
 C-N        &  1  & $+$7.39 &  $-$0.44 & 0.63 & $+$2.21 & 0.63 	&  1  & $+$7.85 &  $+$0.02 & 1.30 & $+$3.04 & 1.06  &  1  & $+$7.49 &  $-$0.34 & 0.50 & $+$1.62 & 0.49  \\ 
\ion{O}{i}  &     &\dots    &  \dots   &\dots &\dots    &\dots  &     & \dots &  \dots &\dots&\dots &\dots&     &\dots  &\dots   &\dots&\dots &\dots\\ 
\ion{Na}{i} &  3  & $+$4.53 &  $-$1.71 & 0.16 & $+$0.95 & 0.18 	&  2  & $+$3.89 &  $-$2.36 & 0.77 & $+$0.66 & 0.69  &  2  & $+$4.62 &  $-$1.63 & 0.39 & $+$0.33 & 0.37  \\ 
\ion{Mg}{i} &  2  & $+$5.12 &  $-$2.49 & 0.39 & $+$0.17 & 0.39 	&  2  & $+$5.36 &  $-$2.25 & 0.88 & $+$0.77 & 0.70  &  3  & $+$6.18 &  $-$1.43 & 0.19 & $+$0.53 & 0.19  \\
\ion{Si}{i} &     & \dots &  \dots &\dots&\dots &\dots&     & \dots &  \dots &\dots&\dots &\dots&     &\dots  &\dots   &\dots&\dots &\dots\\ 
\ion{K}{i}  &  1  & $+$2.98 &  $-$2.05 & 0.32 & $+$0.60 & 0.30 	&  2  & $+$4.18 &  $-$0.86 & 0.72 & $+$2.16 & 0.74  &  1  & $+$3.29 &  $-$1.74 & 0.21 & $+$0.22 & 0.21  \\ 
\ion{Ca}{i} &  3  & $+$4.23 &  $-$2.11 & 0.28 & $+$0.55 & 0.27 	&  5  & $+$4.15 &  $-$2.20 & 0.24 & $+$0.82 & 0.51  &  8  & $+$4.77 &  $-$1.58 & 0.09 & $+$0.38 & 0.11  \\ 
\ion{Sc}{ii} &  3  & $+$0.29 &  $-$2.86 & 0.18 & $-$0.22 & 0.23 	&  2  & $+$0.63 &  $-$2.53 & 0.36 & $+$0.35 & 0.38  &  3  & $+$0.50 &  $-$2.65 & 0.22 & $-$0.63 & 0.23  \\ 
\ion{Ti}{i} &  2  & $+$2.15 &  $-$2.80 & 0.36 & $-$0.15 & 0.35 	&  2  & $+$1.86 &  $-$3.10 & 0.35 & $-$0.08 & 0.53  &  6  & $+$3.00 &  $-$1.95 & 0.14 & $+$0.01 & 0.15  \\ 
\ion{Ti}{ii} &  8  & $+$2.44 &  $-$2.52 & 0.13 & $+$0.12 & 0.15 	&  6  & $+$2.50 &  $-$2.45 & 0.30 & $+$0.42 & 0.40  & 10  & $+$2.81 &  $-$2.14 & 0.11 & $-$0.12 & 0.20  \\ 
\ion{Cr}{i} &  5  & $+$2.74 &  $-$2.90 & 0.16 & $-$0.24 & 0.18 	&     & \dots &  \dots &\dots&\dots &\dots&  8  & $+$3.65 &  $-$1.99 & 0.16 & $-$0.03 & 0.16  \\ 
\ion{Mn}{i} &  2  & $+$2.36 &  $-$3.07 & 0.29 & $-$0.42 & 0.28 	&  2  & $+$2.13 &  $-$3.30 & 0.67 & $-$0.28 & 0.62  &  3  & $+$2.78 &  $-$2.65 & 0.17 & $-$0.69 & 0.18  \\ 
\ion{Fe}{i} & 23  & $+$4.85 &  $-$2.66 & 0.11 & $+$0.00 & 0.00 	& 12  & $+$4.48 &  $-$3.02 & 0.47 & $+$0.00 & 0.00  & 46  & $+$5.54 &  $-$1.96 & 0.08 & $+$0.00 & 0.00  \\ 
\ion{Fe}{ii} &  3  & $+$4.87 &  $-$2.63 & 0.11 & $+$0.00 & 0.00 	&  3  & $+$4.64 &  $-$2.87 & 0.28 & $+$0.00 & 0.00  &  5  & $+$5.48 &  $-$2.02 & 0.21 & $+$0.00 & 0.00  \\ 
\ion{Co}{i} &     & \dots &  \dots &\dots&\dots &\dots&     & \dots &  \dots &\dots&\dots &\dots&  5  & $+$4.23 &  $-$1.99 & 0.14 & $-$0.04 & 0.15  \\ 
\ion{Ni}{i} &  3  & $+$3.49 &  $-$2.73 & 0.38 & $-$0.08 & 0.38 	&     & \dots &  \dots &\dots&\dots &\dots&  2  & $+$2.69 &  $-$2.30 & 0.69 & $-$0.34 & 0.68  \\ 
\ion{Zn}{i} &     & \dots &  \dots &\dots&\dots &\dots&  1  & $+$2.41 &  $-$2.15 & 0.18 & $+$0.86 & 0.46  &     &\dots  &\dots   &\dots&\dots &\dots\\ 
\ion{Sr}{ii} &  1  & $+$0.71 &  $-$2.16 & 0.52 & $+$0.48 & 0.51 	&     & \dots &  \dots &\dots&\dots &\dots&  2  & $+$1.10 &  $-$1.77 & 0.34 & $+$0.25 & 0.33  \\ 
\ion{Y}{ii}  & 10  & $+$0.77 &  $-$1.44 & 0.09 & $+$1.19 & 0.12 	&  3  & $+$0.06 &  $-$2.15 & 0.38 & $+$0.72 & 0.39  & 11  & $-$0.16 &  $-$2.37 & 0.19 & $-$0.35 & 0.18  \\ 
\ion{Zr}{ii} &  5  & $+$1.98 &  $-$0.60 & 0.15 & $+$2.03 & 0.15 	&     & \dots &  \dots &\dots&\dots &\dots&  4  & $+$0.78 &  $-$1.80 & 0.23 & $+$0.22 & 0.23  \\ 
\ion{Ba}{ii} &  3  & $+$1.99 &  $-$0.19 & 0.27 & $+$2.45 & 0.26 	&  3  & $+$1.75 &  $-$0.43 & 0.87 & $+$2.43 & 0.83  &  4  & $+$1.05 &  $-$1.13 & 0.35 & $+$0.90 & 0.26  \\ 
\ion{La}{ii} & 10  & $+$0.48 &  $-$0.62 & 0.06 & $+$2.01 & 0.11 	& 14  & $+$0.53 &  $-$0.57 & 0.16 & $+$2.29 & 0.30  & 15  & $-$0.20 &  $-$1.30 & 0.08 & $+$0.72 & 0.20  \\ 
\ion{Ce}{ii} & 13  & $+$0.92 &  $-$0.66 & 0.09 & $+$1.98 & 0.13 	&  6  & $+$0.53 &  $-$1.05 & 0.32 & $+$1.82 & 0.41  & 15  & $+$0.37 &  $-$1.21 & 0.10 & $+$0.82 & 0.20  \\ 
\ion{Pr}{ii} &  5  & $+$0.00 &  $-$0.72 & 0.15 & $+$1.92 & 0.14 	&  3  & $+$0.18 &  $-$0.55 & 0.40 & $+$2.32 & 0.44  &  5  & $-$0.50 &  $-$1.22 & 0.20 & $+$0.80 & 0.21  \\ 
\ion{Nd}{ii} & 16  & $+$0.89 &  $-$0.53 & 0.10 & $+$2.11 & 0.13 	& 13  & $+$0.65 &  $-$0.77 & 0.15 & $+$2.09 & 0.28  & 13  & $+$0.27 &  $-$1.15 & 0.16 & $+$0.87 & 0.18  \\ 
\ion{Sm}{ii} &  9  & $+$0.14 &  $-$0.82 & 0.10 & $+$1.81 & 0.14 	&  6  & $-$0.19 &  $-$1.15 & 0.17 & $+$1.72 & 0.30  & 10  & $-$0.48 &  $-$1.44 & 0.13 & $+$0.58 & 0.18  \\ 
\ion{Eu}{ii} &  2  & $-$0.70 &  $-$1.22 & 0.46 & $+$1.41 & 0.45 	&  4  & $-$0.16 &  $-$0.68 & 0.18 & $+$2.18 & 0.30  &  2  & $-$1.06 &  $-$1.58 & 0.26 & $+$0.45 & 0.29  \\ 
\ion{Gd}{ii} &  4  & $+$0.34 &  $-$0.73 & 0.11 & $+$1.90 & 0.14 	&  3  & $+$0.29 &  $-$0.78 & 0.71 & $+$2.08 & 0.70  &  2  & $+$0.05 &  $-$1.02 & 0.29 & $+$1.00 & 0.32  \\ 
\ion{Tb}{ii} &     & \dots &  \dots &\dots&\dots &\dots&     & \dots &  \dots &\dots&\dots &\dots&     &\dots  &\dots   &\dots&\dots &\dots\\ 
\ion{Dy}{ii} &  2  & $+$0.37 &  $-$0.73 & 0.69 & $+$1.90 & 0.65 	&     & \dots &  \dots &\dots&\dots &\dots&  1  & $-$0.41 &  $-$1.51 & 0.37 & $+$0.51 & 0.31  \\ 
\ion{Lu}{ii} &  1  & $-$0.61 &  $-$0.71 & 0.28 & $+$1.92 & 0.24 	&  1  & $-$0.01 &  $-$0.11 & 0.42 & $+$2.75 & 0.35  &     &\dots  & \dots  &\dots&\dots &\dots\\ 
\ion{Os}{i} &  1  & $+$0.06 &  $-$1.34 & 0.33 & $+$1.32 & 0.33 	&     & \dots &  \dots &\dots&\dots &\dots&     &\dots  & \dots  &\dots&\dots &\dots\\

\\
\hline
    \end{tabular}}
\end{table*}

\begin{table*}
\caption{Abundance Summary for Stars 125696, 40475, and 97508}
\label{tab:abun2}
    \centering
    \resizebox{\textwidth}{!}{
    \begin{tabular}{lrrrrrrrrrrrrrrrrrr}
    \hline\hline
     & \multicolumn{6}{c}{\textbf{125696}} &\multicolumn{6}{c}{\textbf{40475}} &\multicolumn{6}{c}{\textbf{97508}}\\
 El. & N & $\log\epsilon$ (X) & $\mathrm{[X/H]}$ & $\sigma_{\mathrm{[X/H]}}$ & $\mathrm{[X/Fe]}$ & $\sigma_{\mathrm{[X/Fe]}}$ & N & $\log\epsilon$ (X) & $\mathrm{[X/H]}$ & $\sigma_{\mathrm{[X/H]}}$ & $\mathrm{[X/Fe]}$ & $\sigma_{\mathrm{[X/Fe]}}$ & N & $\log\epsilon$ (X)& $\mathrm{[X/H]}$ & $\sigma_{\mathrm{[X/H]}}$ & $\mathrm{[X/Fe]}$ & $\sigma_{\mathrm{[X/Fe]}}$\\
 &  &  &  & [dex] & & [dex] &  &  &  & [dex] &  & [dex] &  &  &  & [dex] &  & [dex]\\
\hline
%
 C-H        &  1  & $+$7.91 &  $-$0.52 & 0.38 & $+$1.20 & 0.35 &  1  & $+$6.75 &  $-$1.68 & 0.30 & $+$1.11 & 0.28  &  1  & $+$7.12 &  $-$1.31 & 0.57 & $+$0.70 & 0.47  \\  
 C$_2$        &  1  & $+$8.20 &  $-$0.23 & 0.19 & $+$1.49 & 0.18 &  1  & $+$7.14 &  $-$1.29 & 0.21 & $+$1.50 & 0.19  &     & \dots & \dots  &\dots&\dots &\dots\\ 
 C$_{corr}$ &     & $+$8.03 &  $-$0.40 & 0.38 & $+$1.32 & 0.35 &     & $+$7.17 &  $-$1.26 & 0.30 & $+$1.53 & 0.28  &     & $+$7.29 &  $-$1.14 & 0.57 & $+$0.87 & 0.47  \\  
 C-N        &  1  & $+$8.38 &  $+$0.55 & 0.57 & $+$2.27 & 0.54 &  1  & $+$6.96 &  $-$0.87 & 0.49 & $+$1.92 & 0.47  &     & \dots & \dots  &\dots&\dots &\dots\\ 
\ion{O}{i}  &     & \dots & \dots  &\dots&\dots &\dots&  4 & $+$8.02 &  $-$0.67 & 0.11 & $+$2.12 & 0.15  &     & \dots & \dots  &\dots&\dots &\dots\\ 
\ion{Na}{i} &  2  & $+$4.75 &  $-$1.49 & 0.41 & $+$0.23 & 0.38 &  4  & $+$4.56 &  $-$1.68 & 0.18 & $+$1.11 & 0.18  &  2  & $+$4.50 &  $-$1.74 & 0.42 & $+$0.27 & 0.34  \\ 
\ion{Mg}{i} &  4  & $+$5.80 &  $-$1.80 & 0.18 & $-$0.08 & 0.20 &  5  & $+$6.17 &  $-$1.43 & 0.15 & $+$1.36 & 0.15  &  3  & $+$5.43 &  $-$2.17 & 0.24 & $-$0.16 & 0.23  \\
\ion{Si}{i} &     & \dots & \dots  &\dots&\dots &\dots&  2 & $+$5.85 &  $-$1.66 & 0.13 & $+$1.13 & 0.14  &     & \dots & \dots  &\dots&\dots &\dots\\ 
\ion{K}{i}  &  1  & $+$3.87 &  $-$1.16 & 0.25 & $+$0.56 & 0.25 &  1  & $+$2.70 &  $-$2.33 & 0.16 & $+$0.46 & 0.16  &     & \dots & \dots  &\dots&\dots &\dots\\ 
\ion{Ca}{i} & 12  & $+$5.05 &  $-$1.29 & 0.12 & $+$0.43 & 0.16 &  8  & $+$3.82 &  $-$2.53 & 0.14 & $+$0.26 & 0.14  &  6  & $+$4.39 &  $-$1.95 & 0.23 & $+$0.06 & 0.26  \\ 
\ion{Sc}{ii} &  1  & $+$0.76 &  $-$2.39 & 0.24 & $-$0.64 & 0.27 &  4  & $-$0.05 &  $-$3.20 & 0.15 & $-$0.48 & 0.15  &  3  & $+$0.70 &  $-$2.45 & 0.34 & $-$0.39 & 0.36  \\ 
\ion{Ti}{i} &  5  & $+$3.31 &  $-$1.64 & 0.16 & $+$0.08 & 0.20 &  4  & $+$2.42 &  $-$2.53 & 0.27 & $+$0.26 & 0.25  &     & \dots & \dots  &\dots&\dots &\dots\\ 
\ion{Ti}{ii} &  6  & $+$2.98 &  $-$1.97 & 0.12 & $-$0.23 & 0.24 &  7  & $+$2.45 &  $-$2.50 & 0.26 & $+$0.22 & 0.18  &  9  & $+$2.76 &  $-$2.20 & 0.25 & $-$0.13 & 0.20  \\ 
\ion{Cr}{i} &  6  & $+$4.12 &  $-$1.52 & 0.12 & $+$0.20 & 0.17 &  4  & $+$2.72 &  $-$2.92 & 0.22 & $-$0.13 & 0.20  &  4  & $+$3.57 &  $-$2.07 & 0.33 & $-$0.06 & 0.34  \\ 
\ion{Mn}{i} &  4  & $+$2.99 &  $-$2.44 & 0.18 & $-$0.72 & 0.20 &  3  & $+$1.74 &  $-$3.69 & 0.48 & $-$0.91 & 0.46  &     & \dots & \dots  &\dots&\dots &\dots\\ 
\ion{Fe}{i} & 30  & $+$5.78 &  $-$1.72 & 0.12 & $+$0.00 & 0.00 & 52  & $+$4.71 &  $-$2.79 & 0.09 & $+$0.00 & 0.00  & 24  & $+$5.49 &  $-$2.01 & 0.24 & $+$0.00 & 0.00  \\ 
\ion{Fe}{ii} &  2  & $+$5.76 &  $-$1.74 & 0.24 & $+$0.00 & 0.00 &  5  & $+$4.77 &  $-$2.73 & 0.16 & $+$0.00 & 0.00  &  5  & $+$5.44 &  $-$2.06 & 0.21 & $+$0.00 & 0.00  \\ 
\ion{Co}{i} &     & \dots & \dots  &\dots&\dots &\dots&    & \dots & \dots  &\dots&\dots &\dots&     & \dots & \dots  &\dots&\dots &\dots\\ 
\ion{Ni}{i} &     & \dots & \dots  &\dots&\dots &\dots&  2 & $+$3.19 &  $-$3.03 & 0.26 & $-$0.24 & 0.25  &  3  & $+$4.33 &  $-$1.89 & 0.26 & $+$0.11 & 0.24  \\ 
\ion{Zn}{i} &     & \dots & \dots  &\dots&\dots &\dots&    & \dots & \dots  &\dots&\dots &\dots&  1  & $+$2.67 &  $-$1.89 & 0.33 & $+$0.11 & 0.38  \\ 
\ion{Sr}{ii} &     & \dots & \dots  &\dots&\dots &\dots&  1 & $+$0.08 &  $-$2.79 & 0.32 & $-$0.06 & 0.32  &     & \dots & \dots  &\dots&\dots &\dots\\ 
\ion{Y}{ii}  &  6  & $+$0.50 &  $-$1.71 & 0.15 & $+$0.03 & 0.24 &  2  & $-$0.74 &  $-$2.95 & 0.27 & $-$0.22 & 0.23  &  4  & $+$0.70 &  $-$1.51 & 0.32 & $+$0.55 & 0.29  \\ 
\ion{Zr}{ii} &  4  & $+$1.58 &  $-$1.00 & 0.19 & $+$0.75 & 0.23 &     & \dots & \dots  &\dots&\dots &\dots&  3  & $+$0.96 &  $-$1.62 & 0.23 & $+$0.44 & 0.25  \\ 
\ion{Ba}{ii} &  4  & $+$2.07 &  $-$0.11 & 0.26 & $+$1.63 & 0.28 &  4  & $-$1.73 &  $-$3.91 & 0.21 & $-$1.18 & 0.15  &  5  & $+$1.64 &  $-$0.55 & 0.36 & $+$1.52 & 0.26  \\ 
\ion{La}{ii} & 14  & $+$0.60 &  $-$0.50 & 0.10 & $+$1.24 & 0.25 &     & \dots & \dots  &\dots&\dots &\dots&  8  & $+$0.70 &  $-$0.40 & 0.25 & $+$1.66 & 0.23  \\ 
\ion{Ce}{ii} &  8  & $+$0.98 &  $-$0.60 & 0.10 & $+$1.15 & 0.26 &     & \dots & \dots  &\dots&\dots &\dots&  9  & $+$0.78 &  $-$0.80 & 0.15 & $+$1.27 & 0.20  \\ 
\ion{Pr}{ii} &  5  & $+$0.36 &  $-$0.36 & 0.17 & $+$1.39 & 0.28 &     & \dots & \dots  &\dots&\dots &\dots&  3  & $+$0.18 &  $-$0.54 & 0.33 & $+$1.52 & 0.25  \\ 
\ion{Nd}{ii} & 20  & $+$0.85 &  $-$0.57 & 0.13 & $+$1.18 & 0.24 &     & \dots & \dots  &\dots&\dots &\dots&  8  & $+$0.83 &  $-$0.59 & 0.21 & $+$1.47 & 0.21  \\ 
\ion{Sm}{ii} &  9  & $+$0.45 &  $-$0.51 & 0.13 & $+$1.24 & 0.25 &     & \dots & \dots  &\dots&\dots &\dots&  7  & $+$0.29 &  $-$0.67 & 0.29 & $+$1.39 & 0.22  \\ 
\ion{Eu}{ii} &  3  & $-$0.26 &  $-$0.78 & 0.12 & $+$0.96 & 0.24 &     & \dots & \dots  &\dots&\dots &\dots&  2  & $-$0.41 &  $-$0.93 & 0.38 & $+$1.13 & 0.33  \\ 
\ion{Gd}{ii} &  1  & $+$0.74 &  $-$0.33 & 0.29 & $+$1.41 & 0.40 &     & \dots & \dots  &\dots&\dots &\dots&     & \dots & \dots  &\dots&\dots &\dots\\ 
\ion{Tb}{ii} &  1  & $-$0.43 &  $-$0.73 & 0.22 & $+$1.02 & 0.26 &     & \dots & \dots  &\dots&\dots &\dots&     & \dots & \dots  &\dots&\dots &\dots\\ 
\ion{Dy}{ii} &  1  & $+$0.57 &  $-$0.54 & 0.43 & $+$1.21 & 0.37 &     & \dots & \dots  &\dots&\dots &\dots&  1  & $+$0.86 &  $-$0.24 & 0.41 & $+$1.82 & 0.40  \\ 
\ion{Lu}{ii} &     & \dots & \dots  &\dots&\dots &\dots&    & \dots & \dots  &\dots&\dots &\dots&     & \dots & \dots  &\dots&\dots &\dots\\ 
\ion{Os}{i} &     & \dots & \dots  &\dots&\dots &\dots&    & \dots & \dots  &\dots&\dots &\dots&  1  & $+$1.69 &  $+$0.29 & 0.60 & $+$2.30 & 0.57  \\

\\
\hline
    \end{tabular}}
\end{table*}

\begin{table}
\caption{Abundance Uncertainties.}
\label{tab:err}
    \centering
    \resizebox{\columnwidth}{!}{
    \begin{tabular}{llrrrrr}
    \hline\hline
Stellar ID & El & $\Delta_{T_{\rm eff}}$ & $\Delta_{\log g}$ & $\Delta_{\xi}$ & $\Delta_{\mathrm{[M/H]}}$ & $s_X$\\ 
& & [dex] & [dex] & [dex] & [dex] & [dex]\\
\hline
75044 &   C-H        &   0.45 & $-$0.09 &$-$0.09 &   0.10 & 0.00  \\
75044 &   C$_2$        &   0.36 &    0.00 &   0.00 &   0.06 & 0.00  \\
75044 &   C-N        &   0.62 & $-$0.04 &$-$0.02 &   0.22 & 0.00  \\
75044 &  \ion{Na}{i} &   0.15 & $-$0.06 &   0.00 &$-$0.02 & 0.08  \\
75044 &  \ion{Mg}{i} &   0.31 & $-$0.13 &$-$0.07 &$-$0.01 & 0.42  \\
75044 &  \ion{K}{i}  &   0.11 &    0.04 &$-$0.08 &$-$0.18 & 0.00  \\
75044 &  \ion{Ca}{i} &   0.04 & $-$0.00 &$-$0.11 &$-$0.20 & 0.08  \\
75044 &  \ion{Sc}{ii} &$-$0.13 &    0.06 &$-$0.06 &$-$0.09 & 0.10  \\
75044 &  \ion{Ti}{i} &   0.37 & $-$0.06 &$-$0.01 &$-$0.05 & 0.10  \\
75044 &  \ion{Ti}{ii} &$-$0.08 &    0.05 &$-$0.05 &$-$0.00 & 0.26  \\
75044 &  \ion{Cr}{i} &   0.02 & $-$0.03 &$-$0.02 &$-$0.04 & 0.13  \\
75044 &  \ion{Mn}{i} &   0.29 & $-$0.06 &$-$0.03 &$-$0.04 & 0.10  \\
75044 &  \ion{Fe}{i} &   0.07 & $-$0.06 &$-$0.01 &$-$0.02 & 0.12  \\
75044 &  \ion{Fe}{ii} &$-$0.11 &    0.11 &$-$0.06 &   0.05 & 0.01  \\
75044 &  \ion{Ni}{i} &   0.24 & $-$0.02 &$-$0.03 &   0.00 & 0.36  \\
75044 &  \ion{Sr}{ii} &   0.03 &    0.04 &$-$0.14 &   0.02 & 0.00  \\
75044 &  \ion{Y}{ii}  &$-$0.06 &    0.09 &   0.00 &   0.01 & 0.03  \\
75044 &  \ion{Zr}{ii} &$-$0.02 &    0.05 &$-$0.07 &   0.02 & 0.18  \\
75044 &  \ion{Ba}{ii} &   0.17 & $-$0.08 &$-$0.12 &   0.03 & 0.16  \\
75044 &  \ion{La}{ii} &$-$0.09 &    0.10 &$-$0.01 &$-$0.02 & 0.00  \\
75044 &  \ion{Ce}{ii} &$-$0.07 &    0.06 &$-$0.03 &$-$0.01 & 0.15  \\
75044 &  \ion{Pr}{ii} &$-$0.02 &    0.09 &$-$0.02 &   0.04 & 0.16  \\
75044 &  \ion{Nd}{ii} &$-$0.06 &    0.08 &$-$0.02 &   0.01 & 0.13  \\
75044 &  \ion{Sm}{ii} &$-$0.08 &    0.06 &$-$0.04 &$-$0.02 & 0.18  \\
75044 &  \ion{Eu}{ii} &   0.08 &    0.07 &$-$0.02 &   0.07 & 0.43  \\
75044 &  \ion{Gd}{ii} &$-$0.09 &    0.11 &   0.00 &   0.00 & 0.07  \\
75044 &  \ion{Dy}{ii} &   0.42 &    0.03 &$-$0.12 &   0.21 & 0.36  \\
75044 &  \ion{Lu}{ii} &   0.21 &    0.08 &$-$0.00 &   0.12 & 0.00  \\
75044 &  \ion{Os}{i} &   0.10 &    0.07 &$-$0.00 &$-$0.09 & 0.00  \\\\
\hline
    \end{tabular}}
    \tablefoot{The complete version of this Table is available online only. Results for one star is shown here to illustrate its form and content.}  
\end{table}

\subsection{CNO Abundances}
\begin{figure}[hbt!]
\centering
\includegraphics[width=\linewidth]{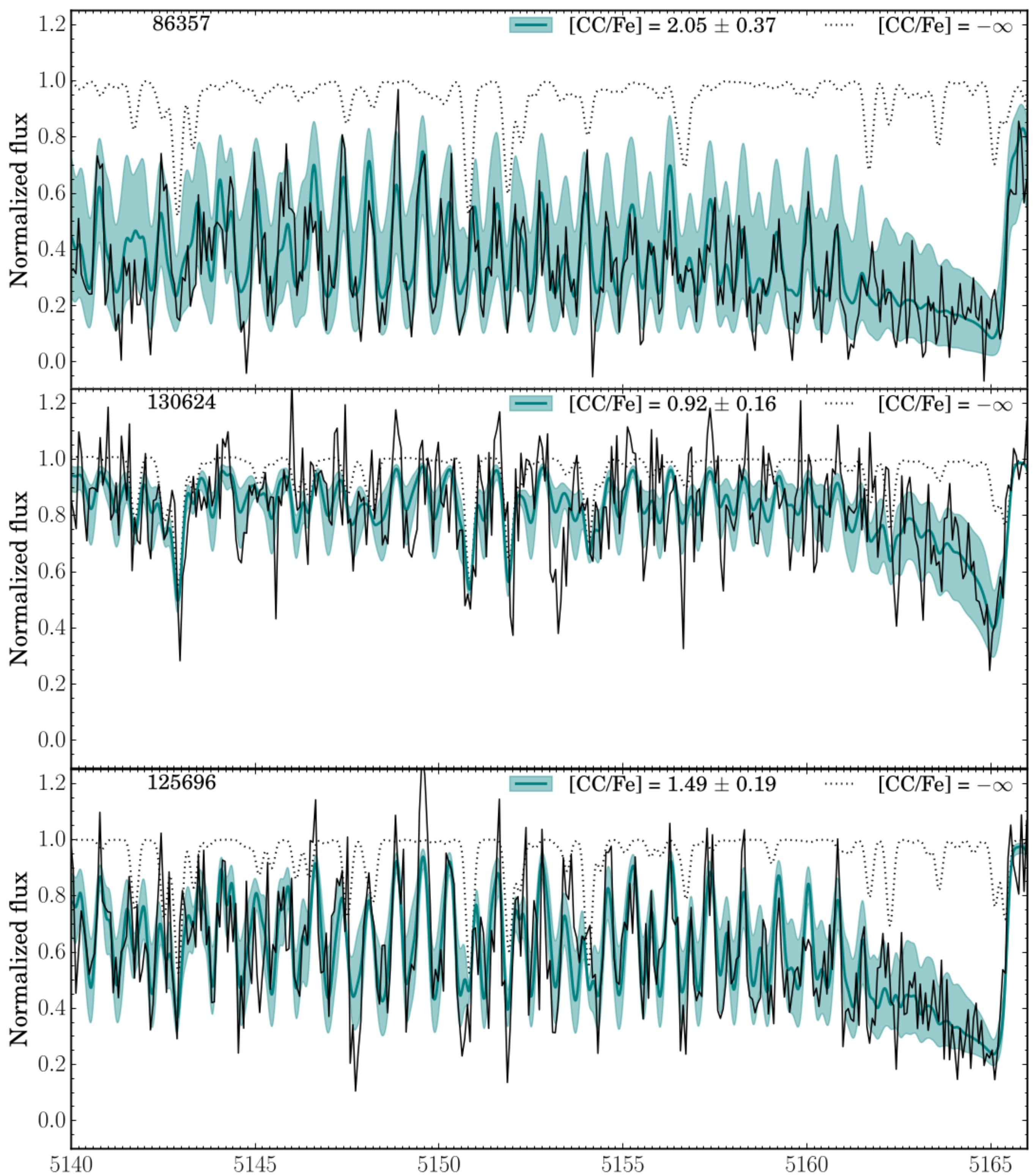}
\caption{\label{fig:synth} Synthesis of the C$_2$ band in 86357 (top), 130624 (middle), and 125696 (bottom). Each panel shows the observed spectrum (solid black line) along with a synthesis without C (dotted line), the best-fit synthesis (solid teal line) and uncertainty on the best-fit synthesis (shaded region).}
\end{figure}

Carbon abundances are derived via synthesis of the C-H $G$-band at 4300\,{\AA} and the C$_2$ swan band at 5160\,{\AA}, when present. In 86357, the C-H band is saturated, and only the C$_2$ band could be used. Generally, the C abundances are in good agreement between the two bands for the stars when both bands are present and not saturated. Synthesis of the C$_2$ band for three of the sample stars with varying C abundances is shown in Figure \ref{fig:synth}. An oxygen abundance could only be derived for one of the stars (40475: $\mathrm{[O/Fe]} = 2.12$) from the 6300\,{\AA} feature. 40475 also exhibit large enhancements in other $\alpha$-elements like Mg and Si, while this is not the case for the other stars in the sample. Thus when deriving C abundances for the rest of the stars, a standard oxygen enhancement for metal-poor stars of $\mathrm{[O/Fe]}=0.4$ was assumed. All of the stars are significantly enhanced in C and, as such, can be labelled CEMP stars. In Table \ref{tab:abun1} and \ref{tab:abun2}, we also list the C abundance corrected for stellar evolutionary effects, C$_{corr}$ and these values are plotted in Figure \ref{fig:tot}. The corrections were made following \citet{placco2014} and using the C-H abundances for all stars, except for 86357, where the C$_2$ abundance was used.  

We also investigated the $^{12}$C/$^{13}$C ratio of the stars by fitting $^{13}$C and $^{12}$C lines in a region around 4230\,{\AA}. We find that using a $^{12}$C/$^{13}$C ratio of 4 gives a good fit to the spectral regions in all of the stars. This ratio is consistent with the equilibrium value for the CNO cycle, suggesting that the material in all of our stars has been well mixed, as is expected for cool red giants. This would equally apply if any of our binary stars experienced mass transfer, as the companion star would also have had a well-mixed outer layer. After determining the C abundances of the stars, N abundances were derived from the C-N band at 4215\,{\AA} for five of the stars (excluding 97508, where the S/N was too low). All five stars exhibit an enhancement in N, which is often also found for halo CEMP stars \citep{hansen2015}. 

\subsection{Light and Iron Peak Element Abundances}
We have derived abundances for Na, Mg, Ca, Sc, Ti, Cr, Mn, and Ni for most of our stars, while a Co abundance could only be derived for one star (130624) and Zn for two stars (97508 and 86357). Our analysis resulted in the first abundances for K derived from any stars in Carina, which displays a spread similar to most other light element abundances in Carina. We were not able to derive abundances for Al or V in any of our stars as the lines were too heavily blended.

\subsection{Neutron-Capture Element Abundances}
In addition to the C enrichment, all but one of the stars (40475) also display large enhancements in neutron-capture elements. In Table \ref{tab:class}, we list the $\mathrm{[Ba/Fe]}$ and $\mathrm{[Ba/Eu]}$ ratios and the corresponding CEMP classification of the stars following \cite{beers2005}. According to this classification, we find one CEMP-no star, three CEMP-$s$ stars, and two CEMP-$r/s$ stars. For the CEMP-no star, abundances of Sr, Y, and Ba are derived, while for the CEMP-$s$ and CEMP-$r/s$ stars, we derive abundance for the neutron-capture elements Y, Zr, Ba, La, Ce, Pr, Nd, Sm, Eu, Gd, and Dy and in some stars, we are also able to derive abundances for Sr, Tb, Lu, and Os.

\begin{table}
\caption{Classification of Program Stars According to \cite{beers2005}.}
\label{tab:class}
    \centering
    \begin{tabular}{lrrc}
    \hline\hline
    Stellar ID & $\mathrm{[Ba/Fe]}$ & $\mathrm{[Ba/Eu]}$ & CEMP type \\
\hline
75044 & 2.45 & 1.04 & CEMP-$s$\\
86357 & 2.43 & 0.61 & CEMP-$s$\\
130624 & 0.90 & 0.45 & CEMP-$r/s$\\
125696 & 1.63 & 0.67 & CEMP-$s$ \\
40475 & $-$1.18& ... & CEMP-no \\
97508 & 1.52 & 0.39 & CEMP-$r/s$\\
\hline
    \end{tabular}
   \end{table}

\section{Discussion \label{sec:discuss}}

\subsection{Comparison to Abundances of Stars in Carina.}
As shown in Figure \ref{fig:tot}, the $\alpha$ and iron-peak element abundances we derive for our six sample stars are similar to those of the CEMP and other Carina stars from the literature. The one outlier is the CEMP-no star in our sample, 40475, which has high Mg and Si abundances and will be discussed below. Abundances for Sr, Ba, and Eu have been derived for a number of stars in Carina. We, therefore, include panels for these elements in Figure \ref{fig:tot}. From this, we see that the Ba abundance of the CEMP-no star 40475 is in good agreement with Ba abundances for other Carina stars at similar metallicity. Thus, from this sample, the behaviour of dSph galaxy CEMP stars follows that of the MW halo CEMP stars, with $\alpha$ and iron-peak element abundances of CEMP-$s$ and $r/s$ stars being similar to other metal-poor stars \citep{norris2013,hansen2015}, while some CEMP-no stars also have distinct light element abundances signatures.

\subsection{CEMP Stars in dSph Galaxies \label{sec:CEMPdsph}}
The first carbon or CH stars in dwarf galaxies were detected more than 40 years ago from low-resolution spectra and in prism surveys. For example, \cite{cannon1981} obtained spectra of two of the brightest members of Carina, finding both to exhibit strong C$_2$ bands, while \citet{mould1982} discovered eight carbon stars in Carina (including the two from \citet{cannon1981}) in their prism survey. Later, \citet{azzopardi1985,azzopardi1986} carried out an extensive prism survey for carbon stars in the dwarf spheroidals Sculptor, Carina, Leo~I, Leo~II, Draco, and Ursa Minor and presented a list of spectroscopically confirmed carbon stars in \citet{azzopardi1986}. High-resolution follow-up for detailed chemical analysis was not possible at the time and is still a challenge due to the faint nature of the stars. However, although the sample presented here constitutes the largest sample of CEMP stars in a dSph for which a detailed abundance analysis has been carried out, a few CEMP stars in other dSph galaxies, including stars discovered by \citet{azzopardi1985}, have also been analysed previously. We have searched the literature for stars with both C, and neutron-capture element abundances reported. After correcting the reported C abundances for evolutionary effects following \citet{placco2014}, we find CEMP-no, $s$, and $r/s$ stars in the following dSph galaxies: Carina, Draco, Sextans, Sagittarius, Sculptor, and Ursa Minor. 

In Draco, the star 19219 analysed by \citet{cohen2009} qualifies as a CEMP star after C correction. These authors derive a Ba abundance of $\mathrm{[Ba/Fe]} = -0.55$. Hence, the star is a CEMP-no star. \citet{cohen2009} reported a radial velocity for the star of $-295.7$\,km\,s$^{-1}$, which is in very good agreement with the recent value of $-296.5\pm2.3$\,km\,s$^{-1}$ reported by \citet{massari2020}, suggesting that the star is not in a close binary system. 

One CEMP star, S15-19, located in Sextans, was analysed by \citep{honda2011}, who found it to be Ba but not Eu enhanced, resulting in a CEMP-$s$ label. They also found evidence of radial velocity variations suggesting that the star belongs to a binary system. One CEMP star was also identified in Sagittarius by \citet{sbordone2020}. The authors found that star to have roughly equal enhancements of Ba and Eu, labelling it a CEMP-$r/s$ star. Three consistent radial velocities are reported for the star by \citet{sbordone2020}, providing no evidence for binary motion.

In Sculptor, four CEMP stars have been analysed. First, a CEMP-no star, ET0097, was discovered by \cite{skuladottir2015}. These authors find that the star has $\alpha$ and iron-peak abundances resembling other Sculptor stars but shows an enhancement in the light neutron-capture elements Sr and Y ($\mathrm{[Sr/Fe] = 0.71}$ and $\mathrm{[Y/Fe]} = 0.34$), leading the authors to suggest that it was polluted by a weak $r$-process. In a later study, the radial-velocity measurements for this star were explored, finding a variation of $> 7$~km$^{-1}$ over a period of nine years, leading the authors to suggest that it is a binary \citep{skuladottir2017}. \citet{salgado2016} analysed the star Scl-1013644 and found high C and Ba abundances, concluding that it is a CEMP-$s$ star. However, the star was previously analysed by \citet{geisler2005}, who derived a similar Ba abundance as \citet{salgado2016}, but also derived an Eu abundance for the star resulting in $\mathrm{[Ba/Eu]} = 0.44$. Hence, the star is actually a CEMP-$r/s$ star. The other two CEMP stars in Sculptor have high Ba abundances and may be CEMP-$s$ stars \citep{lardo2016}. However, no Eu abundances have been derived for the stars; therefore, an $r/s$ classification can not be excluded. None of these stars has multiple radial velocity measurements reported in the literature, rendering their binary nature unknown.  

Three CEMP stars in Ursa Minor have abundances for neutron-capture elements derived. Two stars, UMiJI19 and UMi33533, were analysed by \citet{cohen2010}, who also found the stars to have low Ba abundances ($\mathrm{[Ba/Fe]} = -1.16$ and $-0.99$ respectively), resulting in a CEMP-no classification. Both stars also have consistent radial velocities reported by the \citet{cohen2010} study ($-246.2$ and $-248.9$\,km\,s$^{-1}$, respectively) and the more recent studies by \citet{spencer2018} ($-245.6\pm0.3$ and $-249.4\pm0.3$\,km\,s$^{-1}$) and \citet{pace2020} ($-243.0\pm2.2$ and $-250.2\pm2.2$\,km\,s$^{-1}$) suggesting that they are not binaries. In Ursa Minor, \citet{kirby2015} also found the star Bel60017 to be a CEMP star, and \citet{duggan2018} later derived a Ba abundance of $\mathrm{[Ba/Fe]} = 1.26$ for the star. Hence, the star is either a CEMP-$s$ or $r/s$ star. Three consistent radial velocity measurements are reported in the literature for this star ($-250.0\pm5.9$\,km\,s$^{-1}$; \citealt{armandroff1995}, $-246.2\pm2.1$\,km\,s$^{-1}$; \citealt{kirby2010}, and $-248.4\pm0.6$\,km\,s$^{-1}$; \citealt{spencer2018}), providing no evidence that the star is a binary.	

Finally, abundances from high-resolution spectra have been reported for three CEMP stars in Carina prior to this study. \cite{abia2008} analysed two carbon-rich stars, ALW-6 and ALW-7, identified by \cite{azzopardi1986}. Abundances for the two stars are plotted as cyan circles in Figure \ref{fig:tot}. As can be seen, these stars exhibit very high Ba abundances compatible with the stars being CEMP-$s$ stars. However, no abundance for Eu was derived. Hence, the stars can also be CEMP-$r/s$ stars. Furthermore, the stars were identified as carbon-rich by \cite{azzopardi1986}, and  C/O ratios of 5 and 7 were reported for the stars by \cite{abia2008}, but no $\mathrm{[C/Fe]}$ ratio has been derived for the stars. More recently, high-resolution spectra of two other stars, ALW-1 and ALW-8, from the \cite{azzopardi1986} list were obtained by \cite{susmitha2017}, who analysed ALW-8, finding it to be a CEMP-no star. ALW-1 was not analysed because of its binary nature and low S/N spectrum. Interestingly, the star rejected by \cite{susmitha2017} is included in the sample analysed here, namely 75044. The abundances from \cite{susmitha2017} for the CEMP-no star are plotted as a yellow circle in Figure \ref{fig:tot}. These authors find that the star has $\alpha$ and iron-peak element abundances similar to other Carina stars and similar to the CEMP-no star in Sculptor, they also find it has a small Y overabundance ($\mathrm{[Y/Fe] = 0.29}$). Finally, they do not find evidence of any radial velocity variation for this CEMP-no star \citep{susmitha2017}. 

The results of this literature search combined with the results of the analysis presented here, where we find one CEMP-no, three CEMP-$s$, and two CEMP-$r/s$ stars in Carina, shows us that CEMP stars are indeed present in dSph galaxies. When sample sizes get large enough, CEMP stars of the most common types are indeed detected\footnote{It should be noted that the more rare CEMP-$r$ type has not yet been detected in a dSph galaxy.}. Also, the CEMP stars in dSphs seem to have the same binary properties as the halo CEMP stars. However, CEMP stars, and in particular CEMP-no stars, are predominantly found at low metallicities ($\mathrm{[Fe/H]} < -2$), a range that is still not fully explored in many dSph galaxies. Current samples of stars with multiple abundances derived in dSph galaxies primarily contain stars with $\mathrm{[Fe/H]} > -2.5$  \citep[e.g.][]{norris2017,shetrone2003,hill2019}. In addition, although C abundances can be derived from medium or low-resolution spectra, high-resolution spectra are required to derive abundances for a number of neutron-capture elements, for example, Eu, making it very resource-intensive to gather appropriate samples.

\subsection{Carbon Enhancement as CEMP Classifier}
Studies of CEMP stars in the MW halo have revealed a strong relation between CEMP classifications and the level of carbon enhancement the stars exhibit. This was first noticed by \cite{spite2013}, who plotted the absolute C abundance of CEMP stars as a function of metallicity and discovered two plateaus: A high C plateau, at $A(\rm C) \sim 8.25$ primarily populated by CEMP-$s$ stars and a low C plateau with $A(\rm C) \sim 6.5$ with mostly CEMP-no stars. This work was extended by \cite{yoon2016} to include a larger sample and to also account for the change in C abundance as a result of the evolutionary state of the star \citep{placco2014}, confirming the initial finding.

\begin{figure*}[hbt!]
\centering
\includegraphics[scale=0.5]{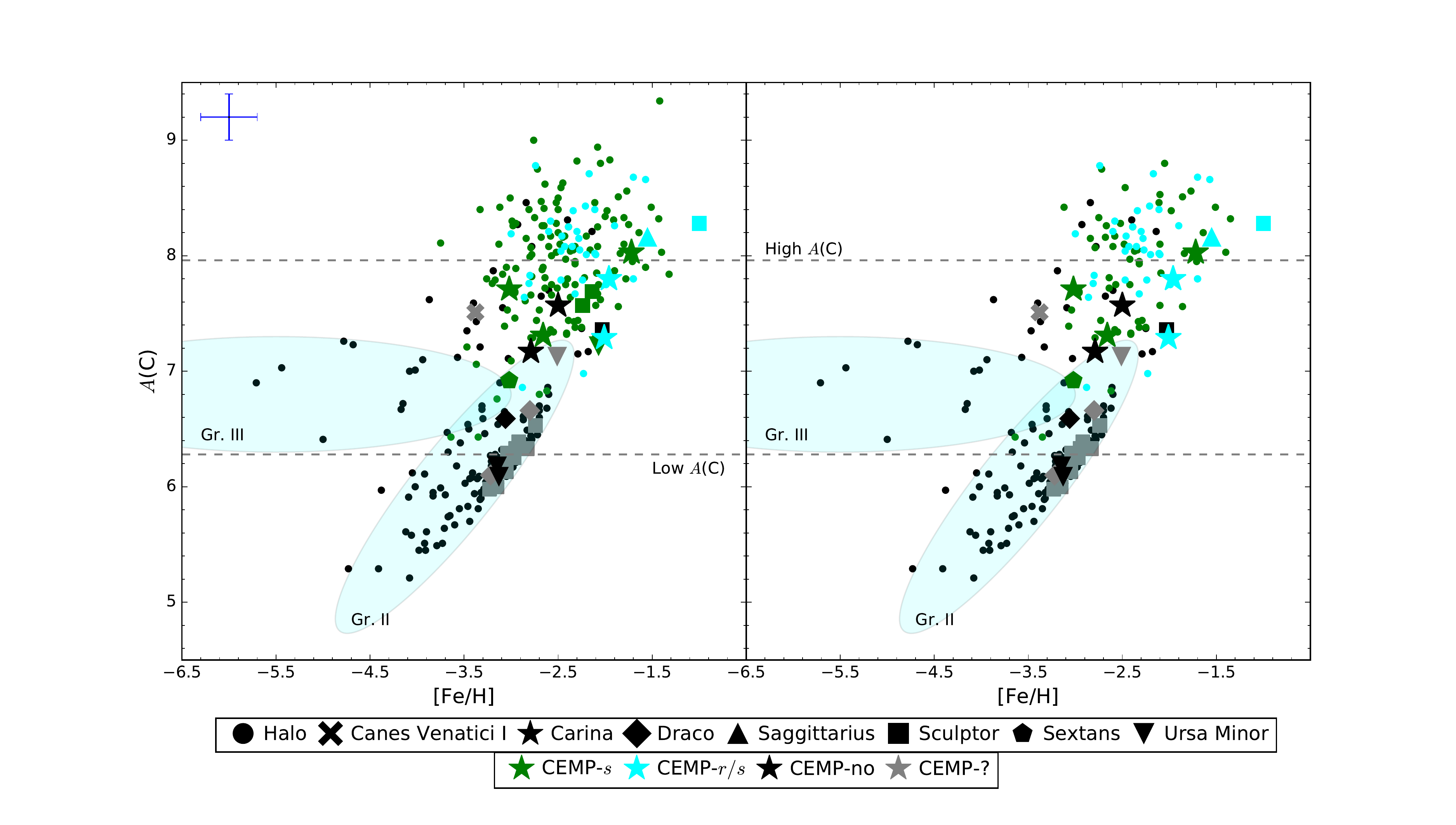}
\caption{Absolute carbon abundance as a function of metallicity for CEMP halo stars (filled circles), along with CEMP stars in the galaxies Canes Venatici~I (cross), Draco (diamonds), Carina (stars), Sculptor (squares), Sextans (downward pointing triangles), and Ursa Minor (upwards pointing triangles). CEMP-no stars are marked in black,  CEMP-$s$ stars in green, CEMP$r/s$ in cyan, and CEMP stars with no classification are marked in grey. Grey dashed lines indicate the high and low $A(\rm C)$ values as found by \citet{yoon2016} while the ellipses mark their groups II and III. The left panel shows stars with high Ba abundances but no Eu abundances as CEMP-$s$ stars, while these are removed from the panel on the right. \label{fig:absC}}
\end{figure*}

In Figure \ref{fig:absC}, we plot the absolute C abundance ($A(\rm C)$) as a function of metallicity for our stars, CEMP stars in the halo \citep{yoon2016,goswami2021,karinkuzhi2021,shejeelammal2021a,shejeelammal2021b,shejeelammal2022}, and dSph CEMP stars from the literature. In addition to the stars listed in the previous section, we also include CEMP stars from dSph galaxies for which no neutron-capture element abundances have been derived. In summary, data from the following dSph galaxies are included; Canes Venatici~I \citep{yoon2020}, Carina \citep{susmitha2017}, Draco \citep{cohen2009,kirby2015,shetrone2013}, Sagittarius \citep{sbordone2020}, Sculptor \citep{skuladottir2015,salgado2016,geisler2005,lardo2016,chiti2018}, Sextans \citep{honda2011}, and Ursa Minor \citep{kirby2015,duggan2018,cohen2010}. All abundances have been converted to the \citet{asplund2009} Solar scale, and C abundances have been corrected for evolutionary effects following \cite{placco2014}. CEMP-no stars are plotted in black symbols, CEMP-$s$ stars in green, CEMP-$r/s$ stars in cyan, and CEMP stars with no classification in grey. The high and low $A(\rm C)$ plateau values determined by \citet{yoon2016} are shown as dashed lines. Looking at Figure \ref{fig:absC}, the $A(\rm C)$ abundances of CEMP stars detected in dSph galaxies follow the trend seen in the halo with high $A(\rm C)$ values found in CEMP-$s$ and CEMP-$r/s$ stars at the higher metallicity range, and lower $A(\rm C)$ values found for the more metal-poor CEMP stars. Confirming the unclassified nature of the low $A(\rm C)$ stars found at low metallicity in e.g., Sculptor as CEMP-no stars would strengthen this picture. However, while $A(\rm C)$ can be used to separate most CEMP-no stars from the bulk CEMP sample, the CEMP-$s$ and $r/s$ stars seem to occupy the same $A(\rm C)$ range, though with the CEMP-$r/s$ stars having more of a preference for $A(\rm C) > 7.5$. 

To construct Figure \ref{fig:absC}, we used the classification criteria from \citet{beers2005}, which require an Eu abundance to distinguish between CEMP-$s$ and CEMP-$r/s$ stars. However, for a large number of stars, no Eu abundance has been derived. This is true for 75 of the 131 stars labelled as CEMP-$s$ stars in \citet{yoon2016}, comprising the majority of the halo sample, and a couple of the stars from the dSph sample (two stars in Sculptor \citep{lardo2016} and one in Ursa Minor \citep{kirby2015,duggan2018}). We have therefore split Figure \ref{fig:absC} into two panels. On the left-hand side, stars with a high Ba abundance but no Eu abundance are included as CEMP-$s$ stars, while these are removed from the right-hand side plot. This exercise shows that the CEMP-$r/s$ stars constitute a significant fraction of the stars with high $A(\rm C)$. In fact, $\sim$40\% of the CEMP stars included in the plot with neutron-capture element enhancements are CEMP-$r/s$ stars. Looking at the plot on the right-hand side of Figure \ref{fig:absC} also suggests that the CEMP-$r/s$ stars generally have higher $A(\rm C)$ values than the CEMP-$s$ stars with a broad transition region, which is likely a signature of their progenitors' surface abundance patterns.

\subsubsection{CEMP-no Stars with High $A(\rm C)$ Values}
From Figure \ref{fig:absC}, it is evident that the CEMP-no stars, both in the halo and in dSph galaxies, occupy a broad region of the  $A(\rm C)$-$\mathrm{[Fe/H]}$ diagram. This led \citet{yoon2016} to separate the CEMP-no stars into two groups; group~II with average low $A(\rm C)$ values and a clear dependence on $\mathrm{[Fe/H]}$ and group~III with higher $A(\rm C)$ values but no dependence on $\mathrm{[Fe/H]}$. These groups are marked with ellipses in Figure \ref{fig:absC} (adapted from \citet{yoon2016}). Following this grouping of the CEMP-no stars, further radial-velocity studies of CEMP stars found that not only do the CEMP-no and CEMP-$s$ stars have different binary properties, but the two groups of CEMP-no stars also seem to have different binary properties. CEMP-no stars with high $A(\rm C)$ are more likely to be binaries than CEMP-no stars with low $A(\rm C)$ \citep{arentsen2019}. The two CEMP-no stars in Carina and the one in Sculptor have high $A(\rm C)$ values, while the CEMP-no stars in Ursa Minor and Draco have lower $A(\rm C)$ values (see Figure \ref{fig:absC}). The CEMP-no star detected in this work, 40475, is likely a binary, as is the CEMP-no star in Sculptor \citep{skuladottir2017}, consistent with their high $A(\rm C)$ values. But radial velocity variations have not been detected for the other Carina CEMP-no star, and the Draco and Ursa Minor CEMP-no stars appear to be single as well (see section \ref{sec:CEMPdsph}) in line with their low $A(\rm C)$ values.

In a recent study \citet{yoon2019} explored the location of dSph CEMP-no stars in the $A(\rm C)$ - $\mathrm{[Fe/H]}$ diagram, and found that the majority of the CEMP-no stars in dSph galaxies fall in group~II. As can be seen in Figure \ref{fig:absC}, this is consistent with our findings. \citet{yoon2019} suggest this is due to the extended star formation and chemical evolution of the dSph galaxies resulting in a strong metallicity dependence for CEMP stars formed in these environments. They also suggest that this can explain the correlation between $A(\rm C)$ and $A(\rm Na)$ and $A(\rm Mg)$ found for group~II halo CEMP-no stars \citep{yoon2016}. Our CEMP-no star exhibits very high Na, Mg, and Si abundances (see Figure \ref{fig:tot}). This signature has been seen in several CEMP-no stars found in the MW halo \citep[][and references therein]{yoon2016} and in ultra-faint dwarf galaxies (Bo\"{o}tes~I:\citet{lai2011,gilmore2013}, Segue~1: \citet{norris2010}, Carina~III:\citet{ji2020a}). However, for the MW stars most of these are group~III CEMP-no stars (with low $A(\rm C)$). The signature has been suggested to be the result of extensive processing and mixing in a massive metal-free spinstar progenitor \citep{maeder2015}. The other dSph CEMP-no stars do not display this signature, although the CEMP-no stars from Ursa Minor and Draco are enhanced in Al \citep{cohen2009,cohen2010}, a signature also seen in some of the most metal-poor halo CEMP-no stars \citep{norris2013}. 

One abundance signature most of the dSph CEMP-no stars do have in common is high $\mathrm{[Sr/Ba]}$ ratios. All except one of the dSph CEMP-no stars listed in Section \ref{sec:CEMPdsph} have $\mathrm{[Sr/Ba]} > 0.8$ (or $\mathrm{[Y/Ba]} > 0.8$, as not all stars have a Sr abundance reported in the literature) \citep{cohen2010,skuladottir2015,susmitha2017}, with the exception being the CEMP-no star in Draco, which has $\mathrm{[Sr/Ba]} = -0.87$ \citep{cohen2009}. This signature has also been discovered in halo CEMP-no stars by \citet{hansen2019}, who suggested CEMP-no stars could be identified based on their Sr to Ba ratio ($\mathrm{[Sr/Ba]} > 0.75$. Following the notion of \citet{yoon2019} that dSph CEMP-no stars mostly fall into group~II, we looked at the $\mathrm{[Sr/Ba]}$ ratios of the group~II CEMP-no stars from \citet{yoon2016} and identified 18 having $\mathrm{[Sr/Ba]} > 0.8$. In the future, studies combining kinematic analysis with results from detailed abundances analysis, as done by \citet{zepeda2022}, will provide important information on the possible accretion history of these stars.

\subsection{Classification of CEMP-$r/s$ Stars}
According to the \cite{beers2005} classification, two of our sample stars, 130624 and 97508, qualify as CEMP-$r/s$ stars, meaning they are potentially enhanced in a mixture of $r$- and $s$-process elements. The origin(s) of this type of abundance pattern of CEMP stars is still not well understood. First of all, it is still debated how these stars should be classified.
The original classification scheme from \cite{beers2005} is based on the Ba and Eu abundances of the stars ($0.0 < \mathrm{[Ba/Eu]} < 0.5$). However, more recently, other classification schemes have been suggested (see \citealt{goswami2021} for a detailed review). 

A classification based on the $\mathrm{[Sr/Ba]}$ ratios of the CEMP stars was put forward by \cite{hansen2019}. Unfortunately, due to the absorption lines being saturated or heavily blended, we are only able to derive Sr abundances for three of our stars, 40475, 75044, and 130624. They have $\mathrm{[Sr/Ba]} = +1.12, -1.97$ and $-0.27$, which results in CEMP-no, CEMP-$r$, and CEMP-$s$ classifications, respectively. Other studies have pointed out that La is a more reliable measure for the $s$-process contribution in these stars as more absorption lines of La than Ba are available in the spectra, and due to the extreme Ba abundances, the Ba lines are often saturated \citep{karinkuzhi2021}. In addition, the Ba abundances also suffer from non-local thermodynamical equilibrium  (non-LTE) effects \citep{korotin2015}. Thus, several studies have suggested using the $\mathrm{[La/Eu]}$ ratios of the stars to separate CEMP-$s$ and $r/s$, where CEMP-$r/s$ stars would have $0.0 < \mathrm{[La/Eu]} < 0.5-0.7$ \citep{frebel2018,goswami2021,karinkuzhi2021} (the upper limit varies depending on the study). \citet{goswami2021} also found that most CEMP-$r/s$ stars have $\mathrm{[Eu/Fe]} > 1$ while CEMP-$s$ stars have $\mathrm{[Eu/Fe]} < 0$. 

Finally, \cite{karinkuzhi2021} investigated a classification method based on the distance of the stellar neutron-capture element abundance pattern from the Solar system $r$-process abundance pattern. Stars enhanced in pure $r$-process material are known to exhibit a quite robust abundance pattern for the heavy neutron-capture elements matching that of the scaled Solar system $r$-process abundance pattern \citep{sneden2008}, so in principle, the more $r$-process dominated a star is, the closer its neutron-capture element abundance pattern will be to the scaled Solar system $r$-process abundance pattern. \cite{karinkuzhi2021} calculates a distance to the scaled Solar $r$-process pattern, $d$, and finds the CEMP-$r/s$ stars in their sample to have $0.5 \leq d \leq 0.8$ while their CEMP-$s$ stars all have $d > 0.7$. We have determined $\mathrm{[La/Eu]}$ values and calculated $d$ for our stars. The results and re-classifications are listed in Table \ref{tab:class_new}. When calculating $d$, we exclude the light neutron-capture elements Sr, Y, and Zr as the abundances of these elements in $r$-process enhanced stars have been found to display a larger scatter than abundance for elements heavier than Ba \citep{hansen2012,ji2016} with respect to, e.g. Eu. For the stars in our sample, these two methods agree on the classification, and we end up with three CEMP-$r/s$ stars in our sample (86357, 130624, and 125696) and two CEMP-$s$ stars (75044 and 97508), which is slightly different from the \cite{beers2005} classification and highlights a likely notable contribution to the abundance patterns of these stars from a nucleosynthesis process with a higher neutron flux than the $s$-process. 

The uncertainties on the derived abundances, of course, present a limitation to the classification. We have estimated this uncertainty by generating 100 sets of "new" abundances for the stars drawn from a Gaussian distribution in the range $\mathrm{[X/Fe]}\pm\sigma_\mathrm{[X/Fe]}$ and re-calculating $d$ with these "new" abundances. The result of this exercise is that $d$ values for the three CEMP-$r/s$ fluctuate in the range from 0.2 to 0.7 while the $d$ values for the two CEMP-$s$ stars lie in the range from 0.2 to 0.9. Thus while the $d$ values vary a great deal, the CEMP-$r/s$ stars still stay below (or just at) the $d = 0.7$ found to be the lower limit for CEMP-$s$ stars in \citet{karinkuzhi2021}.

\begin{table}
\caption{Classification of Program Stars following \cite{karinkuzhi2021}.}
\label{tab:class_new}
    \centering
    \begin{tabular}{lrrc}
    \hline\hline
    Stellar ID & $d$  & $\mathrm{[La/Eu]}$ & CEMP type \\
\hline
75044 & 0.72 & 0.60 & CEMP-$s$\\
86357 & 0.30 & 0.11 & CEMP-$r/s$\\
130624 & 0.59 & 0.27 & CEMP-$r/s$\\
125696 & 0.54 & 0.28 & CEMP-$r/s$ \\
97508 & 0.73 & 0.53 & CEMP-$s$\\
\hline
    \end{tabular}
   \end{table}

\subsection{Origin of CEMP-$r/s$ Stars}
Many attempts have been made to understand the origin(s) of the CEMP-$r/s$ stars' peculiar abundance pattern, steadily assuming that both the $s$- and the $r$-process indeed contribute to these unusual patterns. Most of the invoked scenarios involve various combinations of $r$- and $s$-process events polluting the star or the gas it formed from. For example, forming a binary system from gas already polluted by an $r$-process event and then invoking mass transfer in the binary system to get the additional C and $s$-process enrichment (see \citealt{jonsell2006} and \citealt{abate2016} for reviews). \citet{sbordone2020} found that the latter scenario best explained the abundance pattern of the CEMP-$r/s$ star in the Sagittarius dSph. However, so far, it has only been possible to match a combined $r$- and $s$-process pattern to the abundance pattern of one CEMP-$r/s$ star in the MW halo \citep{gull2018}, which the authors suggest to re-label as CEMP-$r+s$ to reflect the actual contributions of both of these processes. Hence in recent years, the suggestion that a process with a neutron flux intermediate between the $r$- and $s$-process, the so-called $i$-process \citep{cowan1977} could create this signature has gained ground (e.g., \citealt{hampel2016}). 

\begin{figure}
\centering
\includegraphics[scale=0.42]{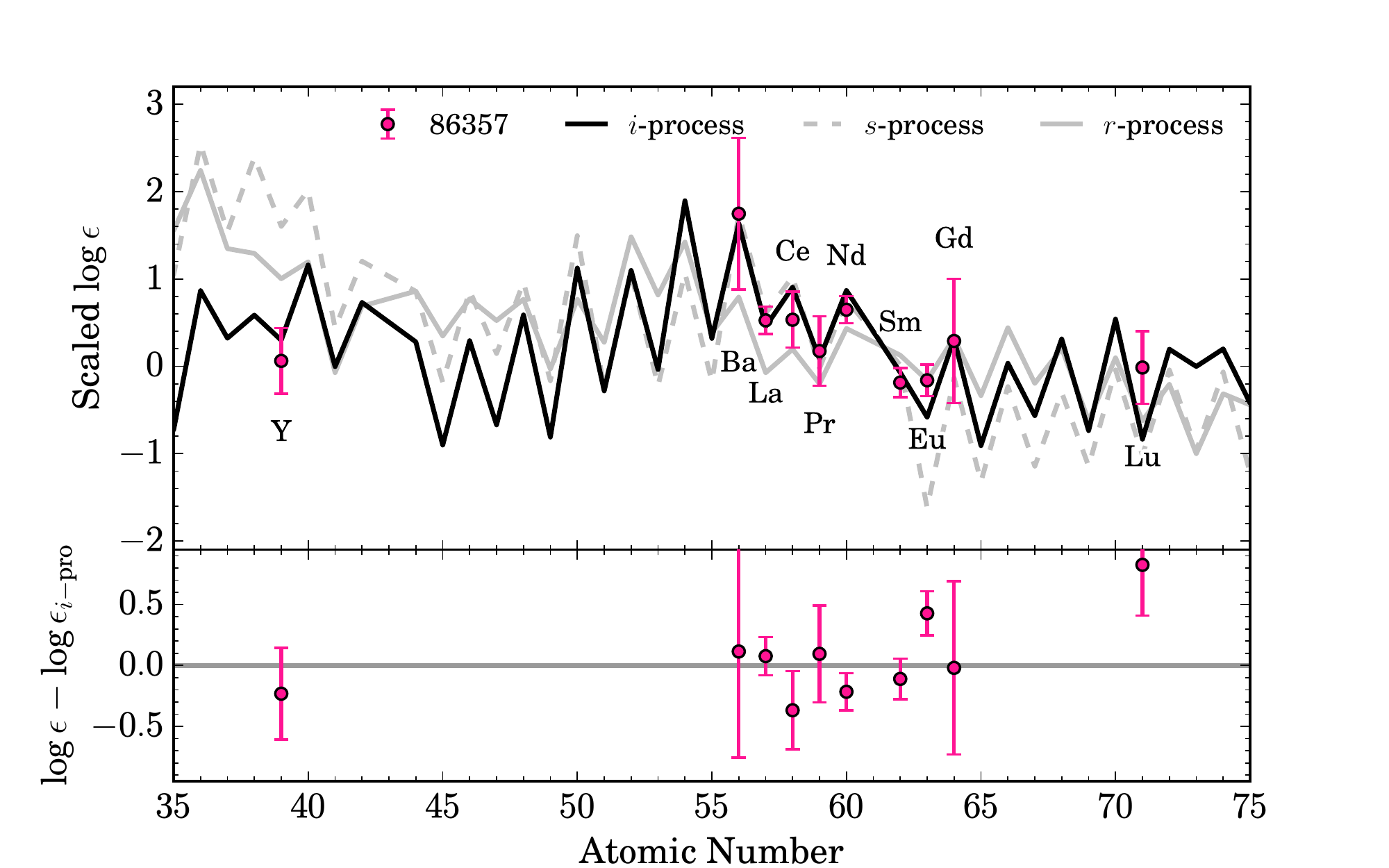}
\includegraphics[scale=0.42]{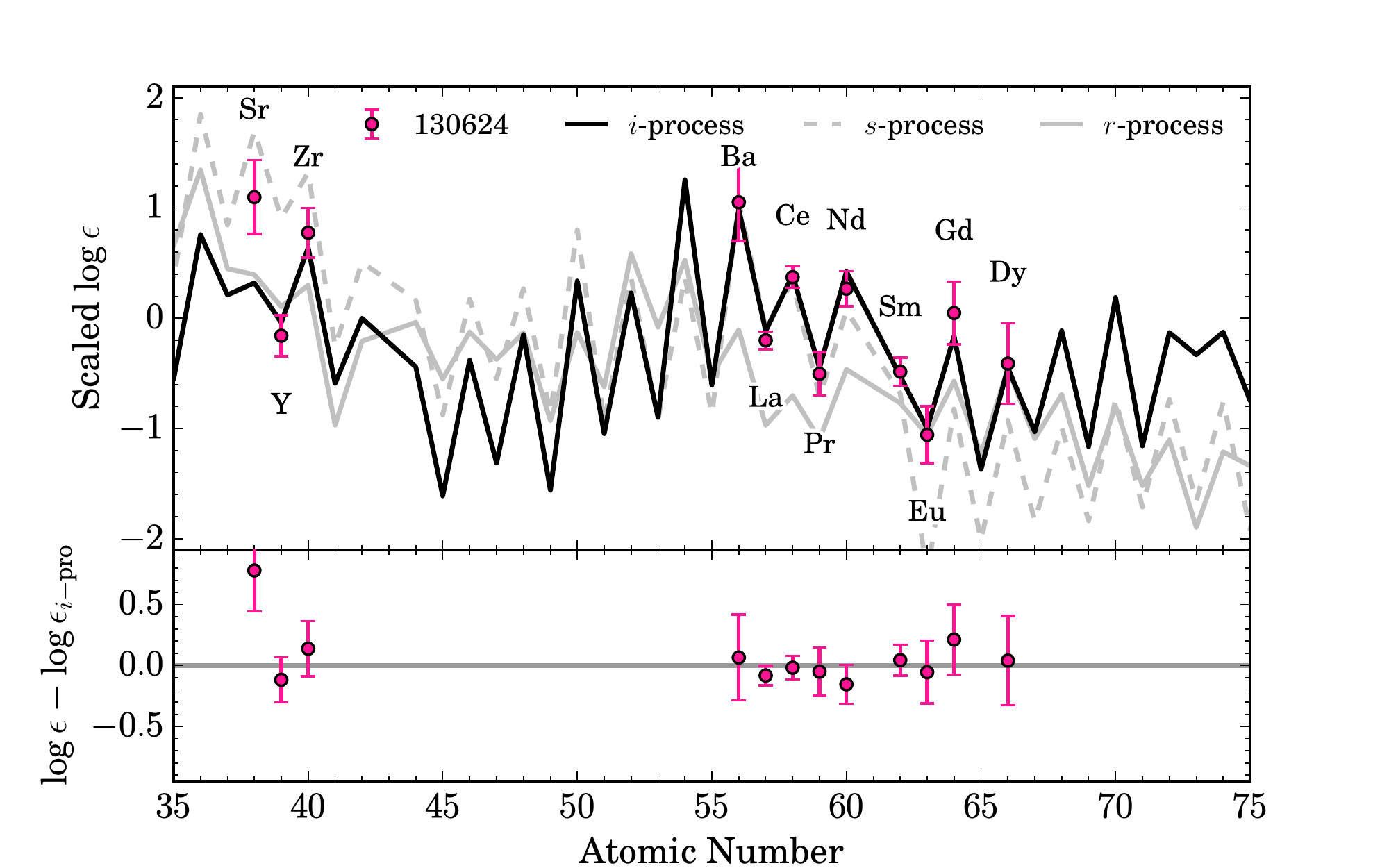}
\includegraphics[scale=0.42]{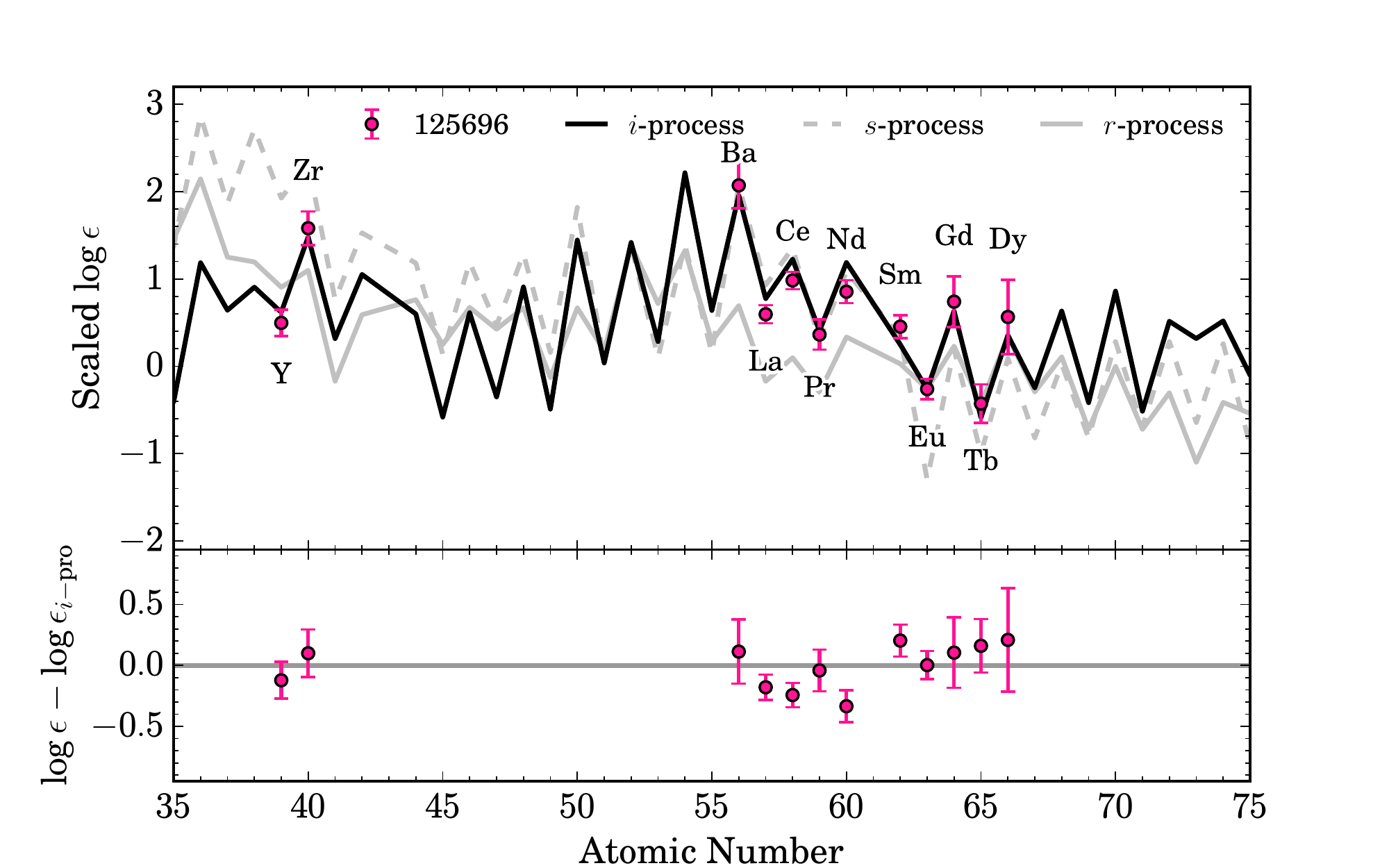}
\caption{Absolute $\log \epsilon$ (X) abundances for the three CEMP-$r/s$ stars in our sample compared to scaled $i$-process yields from \citet{choplin2022} and the scaled Solar $r$- and $s$-process patterns from \cite{sneden2008}. The bottom panel in each plot shows the residual for the match to the $i$-process yields. The Solar $s$- and $r$-process patterns have been scaled to the stellar Ba and Eu abundances, respectively, while the average difference between yields and stellar abundances from Ba to Dy has been used to scale the $i$-process yields.\label{fig:pattern}}
\end{figure}

Currently, various processes in very metal-poor, low mass AGB stars \citep{campbell2008,cristallo2009,campbell2010,stancliffe2011,choplin2021} and super-AGB stars \citep{doherty2015,jones2016} are suggested as $i$-process sources. In addition, conditions required for the $i$-process have also been hypothesized in accreting white dwarfs \citep{denissenkov2019} and during the He shell burning phase of metal-free massive stars \citep{banerjee2018}. Some of these suggested $i$-process sites imply that the CEMP-$r/s$ stars were polluted via mass transfer in a binary system, while others do not. No dedicated radial-velocity monitoring campaign has been published for the CEMP-$r/s$ stars. However, of the three CEMP-$r/s$ stars included in \citet{hansen2016b}, all were found to exhibit radial-velocity variations consistent with belonging to binary systems. For the CEMP-$r/s$ stars detected here, two (125696 and 130624) show clear radial-velocity variations. Thus, it is likely that the observed signature is a result of binary mass transfer similar to that of the CEMP-$s$ signature. This is largely in good agreement with $i$-process conditions being found in specific AGB stars, as mentioned above. In addition, \citet{hampel2016} compared the abundance patterns of 20 CEMP-$r/s$ stars to $i$-process model yields from thermally pulsing AGB stars. They found that the $i$-process yields successfully reproduced the neutron/capture abundance pattern of the CEMP-$r/s$ stars, further supporting this scenario. This study, however, did not include C abundances as a constraint. In the future, it would be interesting to see if the proposed sites of the $i$-process also can produce the high C enhancement seen in CEMP-$r/s$ stars.

In Figure \ref{fig:pattern}, we compare the neutron-capture element abundance patterns of our CEMP-$r/s$ stars to the most recent $i$-process yield available in the literature from \citet{choplin2022} and the Solar $r$- and $s$-process abundance patterns from \cite{sneden2008}. The Solar $s$- and $r$-process patterns have been scaled to the stellar Ba and Eu abundances, respectively, while the average difference between yields and stellar abundances from Ba to Dy has been used to scale the $i$-process yields. The best fits were obtained using the yields from model
M1.0z2.3 with initial mass 1~M$_\odot$ and metallicity $\mathrm{[Fe/H]} = -2.3$ for the stars 86357 and 125696, and yields from model M1.0z2.5 with initial mass 1~M$_\odot$ and metallicity $\mathrm{[Fe/H]} = -2.5$ for the star 130624. For all three stars, the $i$-process model yields are generally a good match to the stellar abundances in the element range from Ba to Dy, while some larger discrepancies are seen for the lighter elements Sr, Y, and Zr and for Lu in 86357. In comparison, neither the Solar $s-$ nor $r$-process patterns display a good match to the stellar abundances. \citet{choplin2022} also provides yields for C, which for the two models used in Figure \ref{fig:pattern} is $\log_\epsilon$(C) = 9.26 (M1.0z2.5) and 9.29 (M1.0z2.3). Scaling these using the same scaling as for the neutron-capture elements, we find $\log_\epsilon$(C) = 7.24, 7.56, and 7.65 for 86357, 12569, and 130624, respectively, which are slightly lower than but almost within uncertainties of the C abundances we derive for the stars ($\log_\epsilon$(C$_{corr}$) = 7.71$\pm0.49$, 8.03$\pm0.38$, and 7.80$\pm0.35$, respectively).

\section{Summary} \label{sec:summary}
We present detailed abundances for six CEMP stars in the Carina dSph galaxy. This is the largest sample of CEMP stars in any dSph for which detailed abundances have been derived. Our sample is comprised of one CEMP-no star, three (two) CEMP-$s$, and two (three) CEMP-$r/s$ stars (depending on the classification criteria). 

We find that the CEMP stars have similar $\alpha$ and iron-peak element abundances as the carbon-normal Carina stars, except for the CEMP-no star, which is highly enhanced in Na, Mg, and Si, a feature also seen in some halo CEMP-no stars. Plotting the absolute C abundance of CEMP stars in dSph galaxies as a function of $\mathrm{[Fe/H]}$ shows a similar trend as what is seen for halo CEMP stars with two $A(\rm C)$ groups, where the CEMP-$s$ and $r/s$ stars mostly occupy the higher plateau and the CEMP-no stars the lower plateau. We find, however, that the CEMP-$r/s$ stars primarily occupy the highest $A(\rm C)$ range with $A(\rm C) > 7.5$ though with a notable overlap region between the CEMP-$s$ and $r/s$ stars. The considerable overlap in $A(\rm C)$ values for these two types of CEMP stars highlights the need for neutron-capture element abundances being derived to identify the CEMP-$r/s$ stars, which will lead to a better understanding of this abundances signature. 

Finally, we match the neutron-capture element abundance pattern and C abundances of our CEMP-$r/s$ stars (\cite{karinkuzhi2021} classification) to $i$-process yield from metal-poor low mass AGB models \citep{choplin2022}, finding a good match for the elements in the range from Ba to Dy, which strengthens the hypothesis that CEMP-$r/s$ stars show the signature of an intermediate neutron-capture process. The models, however, have some trouble reproducing the observed abundances for the light elements and also produce slightly less C compared to the stellar abundances. Hence some work is still needed to understand the nature of the full abundance pattern of these stars.

\begin{acknowledgements}
The authors thank the anonymous referee for their comments, which have improved the manuscript
We thank Josh Adams for his work on the MagE observing program from which the targets of this study were selected. The authors thank Asa Skuladottir for useful discussions on CEMP stars in dwarf galaxies. 
TTH acknowledges support from the Swedish Research Council (VR 2021-05556).  JDS and TTH were partially supported by the NSF through grant AST-1714873.
AF acknowledges support from  NSF CAREER grant AST-1255160 and NSF grant AST-1716251. 
TSL acknowledges financial support from Natural Sciences and Engineering Research Council of Canada (NSERC) through grant RGPIN-2022-04794.

This paper includes data gathered with the 6.5 meter Magellan Telescopes located at Las Campanas Observatory, Chile.
This research made extensive use of the SIMBAD database operated at CDS, Strasbourg, France \citep{wenger2000}, \href{https://arxiv.org/}{arXiv.org}, and NASA's Astrophysics Data System for bibliographic information.
\end{acknowledgements}


\bibliographystyle{aa}
\bibliography{Terese}

\begin{appendix} 
\section{Photometry}

\begin{table}[ht]
\caption{Photometry and derived $T_{\rm eff}$ for the six sample stars.}
\label{tab:photo}
    \centering
    \begin{tabular}{lrrrrrccccc}
    \hline\hline
    Stellar ID & $V$ & $I$ & $J$ & $H$ & $K$ & $T_{\rm eff}(V-I)$ & $T_{\rm eff}(V-J)$ & $T_{\rm eff}(V-H)$ & $T_{\rm eff}(V-K)$ & $T_{\rm eff}(J-K)$\\
   \hline
75044 & 17.708 & 16.284 & 15.235  & 14.557  & 14.295  &4139&4140& \dots& \dots& \dots\\ 
86357 & 18.524 & 17.235 & 16.270  & 15.630  & 15.510  &4332&4347&4238&4308&4419\\  
130624 & 18.629 & 17.476 & 16.551  & 16.039  & \dots &4672&4595&4599& \dots & \dots\\  
125696 & 18.814 & 17.747 & \dots & \dots & \dots &4868& \dots & \dots & \dots& \dots\\ 
40475 & 19.054 & 17.885 & \dots & \dots & \dots &4546& \dots & \dots & \dots& \dots\\ 
97508 & 19.013 & 17.965 & \dots & \dots & \dots &4866& \dots & \dots & \dots& \dots \\ 
\hline
    \end{tabular}
    \tablebib{ $V$ and $I$ magnitudes are from \citet{fabrizio2016} and 2MASS $JHK$ magnitudes were taken from \citet{cutri2003}}
   \end{table}

\section{Atomic data}
\begin{table}[ht]
\caption{Atomic Data and Abundances for Individual Lines Analysed.}
\label{tab:lines}
    \centering
    \resizebox{\columnwidth}{!}{
    \begin{tabular}{lrrrrrrrr}
    \hline\hline
Stellar ID & Species & $\lambda$ & $\chi$ & $\log$ $gf$ & EW & $\sigma_{\rm EW}$ & $\log \epsilon (X)$ & Ref\\
 &  & (\AA) & (eV) &  & (m\AA) & (m\AA) & (dex) & \\    
\hline
75044 & 11.0 & 5688.20 & 2.10 & $-$0.41 & 53.18  & 6.77  & 4.46 & 1\\
75044 & 11.0 & 5889.95 & 0.00 & $+$0.11 & 481.72 & 10.48 & 4.66 & 1\\
75044 & 11.0 & 5895.92 & 0.00 & $-$0.19 & 368.27 & 8.08  & 4.52 & 1\\
75044 & 12.0 & 5172.68 & 2.71 & $-$0.36 & 246.40 & 12.37 & 4.60 & 2\\
75044 & 12.0 & 5183.60 & 2.72 & $-$0.17 & 455.35 & 13.94 & 5.43 & 2\\
75044 & 19.0 & 7698.96 & 0.00 & $-$0.18 & synth  & synth & 2.98 & 1\\
75044 & 20.0 & 6102.72 & 1.88 & $-$0.81 & synth  & synth & 4.07 & 3\\
75044 & 20.0 & 6122.22 & 1.89 & $-$0.33 & synth  & synth & 4.25 & 3\\
75044 & 20.0 & 6161.29 & 2.52 & $-$1.35 & synth  & synth & 4.11 & 3\\
75044 & 21.1 & 4324.99 & 0.60 & $-$0.44 & synth  & synth & 0.23 & 4\\
75044 & 21.1 & 4374.47 & 0.62 & $-$0.46 & synth  & synth & 0.33 & 4\\
75044 & 21.1 & 4400.39 & 0.61 & $-$0.54 & synth  & synth & 0.29 & 4\\
75044 & 22.0 & 4840.87 & 0.90 & $-$0.43 & 45.81  & 5.04  & 2.20 & 5\\
75044 & 22.0 & 5210.38 & 0.05 & $-$0.82 & synth  & synth & 1.89 & 5\\
75044 & 22.1 & 4443.80 & 1.08 & $-$0.71 & synth  & synth & 2.44 & 6\\
75044 & 22.1 & 4450.48 & 1.08 & $-$1.52 & synth  & synth & 2.15 & 6\\
75044 & 22.1 & 4501.27 & 1.12 & $-$0.77 & synth  & synth & 2.89 & 6\\
75044 & 22.1 & 4571.97 & 1.57 & $-$0.31 & synth  & synth & 2.05 & 6\\
75044 & 22.1 & 5226.54 & 1.57 & $-$1.26 & synth  & synth & 2.61 & 7\\
75044 & 22.1 & 5336.79 & 1.58 & $-$1.60 & synth  & synth & 2.76 & 6\\\\
\hline
    \end{tabular}}
     \tablebib{(1) \citet{kramida2018}; (2) \citet{pehlivan2017}; (3) \citet{yu2018}; (4) \citet{lawler1989}, using hfs from \citet{kurucz1995}; (5) \citet{lawler2013}; (6) \citet{wood2013}; (7) \citet{pickering2001}, with corrections given in \citet{pickering2002}; (8) \citet{sobeck2007}; (9) \citet{denhartog2011} for both log(gf) value and hfs; (10) \citet{obrian1991}; (11) \citet{denhartog2014}; (12) \citet{belmonte2017}; (13) \citet{ruffoni2014}; (14) \citet{denhartog2019}; (15) \citet{melendez2009}; (16) \citet{lawler2015} for log(gf) values and HFS; (17) \citet{wood2014b}; (18) \citet{roederer2012}; (19) \citet{biemont2011}; (20) \citet{ljung2006}; (21) \citet{kramida2018}, using HFS/IS from \citet{mcwilliam1998} when available; (22) \citet{lawler2001a}, using HFS from \citet{ivans2006}; (23) \citet{lawler2009};  (24) \citet{li2007}, using HFS from \citet{sneden2009}; (25) \citep{ivarsson2001}, using HFS from \citet{sneden2009}; (26) \citet{denhartog2003}, using HFS/IS from \citet{roederer2008}; (27) \citet{lawler2006}, using HFS/IS from \citet{roederer2008}; (28) \citet{lawler2001c}, using HFS/IS from \citet{ivans2006}; (29) \citet{denhartog2006}; (30) \citet{lawler2001b}, using HFS from \citet{lawler2001d}; (31) \citet{wickliffe2000}; (32) \citet{lawler2009} for $\log gf$ values and HFS; (33) \citet{quinet2006}
.}
\tablefoot{The complete version of this Table is available online only. A subset is shown here to illustrate its form and content.}    
\end{table}

\end{appendix}
\end{document}